\documentclass{article}
\usepackage{graphics,epsfig,amsfonts,amsmath}
\usepackage{placeins}
\begin{document}
\newtheorem{theorem}{Theorem}[section]
\newtheorem{lemma}[theorem]{Lemma}
\newtheorem{proposition}[theorem]{Proposition}
\newtheorem{corollary}[theorem]{Corollary}
\newenvironment{proof}[1][Proof]{\begin{trivlist}
\hskip \labelsep {\bfseries #1}]}{\end{trivlist}}
\newenvironment{definition}[1][Definition]{\begin{trivlist}
\item[\hskip \labelsep {\bfseries #1}]}{\end{trivlist}}
\newenvironment{example}[1][Example]{\begin{trivlist}
\item[\hskip \labelsep {\bfseries #1}]}{\end{trivlist}}
\newenvironment{remark}[1][Remark]{\begin{trivlist}
\item[\hskip \labelsep {\bfseries #1}]}{\end{trivlist}}
\newcommand{\qed}{\nobreak \ifvmode \relax \else
      \ifdim\lastskip<1.5em \hskip-\lastskip
      \hskip1.5em plus0em minus0.5em \fi \nobreak
      \vrule height0.75em width0.5em depth0.25em\fi}
\newcommand{\N}{\mathbb{N}}
\newcommand{\R}{\mathbb{R}}
\newcommand{\Q}{\mathbb{Q}}
\title{A general framework for modeling tumor-immune system competition and immunotherapy: mathematical analysis and biomedical inferences.}
\author{Alberto d'Onofrio $^{1}$
} % IMPORTANT: leave this curly bracket as the first character of this line.

% Date - leave this blank.
\date{ \today }
\maketitle
%
% Institutions
% Here fill in ur institute name(s) and address(es)
% The number in $^...$ indicates the author number.  For example
{\center $^1$ Department of Epidemiology and Biostatistics,
 European Institute of Oncology, Via Ripamonti 435, Milano, Italy,
 I-20141\\EMAIL:
alberto.donofrio@ieo.eu, PHONE:+390257489819\\[3mm]  }
\begin{abstract}
In this work we propose and investigate a family of models, which admits as particular cases some well known mathematical models of tumor-immune system interaction, with the additional assumption that the influx of immune system cells may be a function of the number of cancer cells. Constant, periodic and impulsive therapies (as well as the non-perturbed system) are investigated both analytically for the general family and, by using the model by Kuznetsov et al. (V. A. Kuznetsov, I. A. Makalkin, M. A. Taylor and A. S. Perelson. Nonlinear dynamics of immunogenic tumors: Parameter estimation and global bifurcation analysis. {\em Bulletin of
Mathematical Biology},{\bf 56(2)} 295-321, (1994)), via numerical simulations.  Simulations seem to show that the shape of the function modeling the therapy is a crucial factor only for very high values of the therapy period $T$, whereas for realistic values of $T$, the eradication of the cancer cells depends on the mean values of the therapy term.
Finally, some medical inferences are proposed.
\end{abstract}
{\bf Keywords: Cancer -- Immunotherapy -- Stability Theory -- Periodical forcing}

\section{Introduction}
Millions of people die from cancer every year \cite{[DBTV]}.
And worldwide trends indicate that millions more will die from this disease in the future \cite{[QDB]}. Great progress has been achieved in fields of cancer prevention and surgery and many novel drugs are
available for medical therapies \cite{[CancerCode],[PPV],[VHR]}.
Biophysical models may prove to be useful in oncology not only in
explaining basic phenomena \cite{[BMB],[W]}, but also in helping
clinicians to better and more scientifically plan the schedules of
the therapies \cite{[W],[mbs]}. An interesting therapeutic
approach is immunotherapy \cite{[PPV],[VHR]}, consisting in
stimulating the immune system in order to better fight, and hopefully eradicate, a cancer. In particular, in this paper I will be
referring to generic immunostimulations, for example via cytokines,
but for the sake of simplicity I will use the term "immunotherapy".
The basic idea of immunotherapy is simple and promising, but the
results obtained in medical investigations are globally
controversial \cite{[Seminar],[Euronc],[Marras],[Kaminski]}, even
if in recent years there has been evident progress. From a
theoretical point of view, a large body of research has been
devoted to mathematical models of cancer-immune system
interactions  and to possible applications to cure the disease
\cite{[Stepanova],[Kuznetsov],[Kirschner],[Ortega],[Bellomo],[NaniMbs],[Galach],[Suzanna],[Sotolongo],[deVladar],[Bellomo2004]}
(and references therein). Analyzing the best known finite
dimensional models
\cite{[Stepanova],[Kuznetsov],[Kirschner],[Galach],[deVladar]}, we
note that their main features are the following:

\begin{itemize}
    \item Existence of a tumor free equilibrium;
    \item Depending on the values of parameters, there is the possibility that the tumor size may tend to $+\infty$ or to a macroscopic value;
    \item Possible existence of a "small tumor size" equilibrium, which coexists with the tumor free equilibrium.
\end{itemize}

An "accessory" feature is the existence of limit cycles
\cite{[Kirschner]}. From this rough summary, one may understand
that the puzzling results obtained up to now by immunotherapy
\cite{[Seminar]} may be strictly linked to the complex dynamical
properties of the immune system-tumor competition. In general, it
happens that the cancer-free equilibrium coexists with other
stable equilibria or with unbounded growth, so that the success of
the cure depends on the initial conditions, and - even
theoretically - it is not always granted.

\section{A general family of models and its properties}

In \cite{[Sotolongo]}, Sotolongo-Costa et al. proposed the
following very interesting Volterra-like model (similar to the one
in \cite{[Galach]}) for the interaction between a population of
tumor cells (whose number is denoted by $X$) and a population of
lymphocyte cells ($Y$) :
\begin{eqnarray}
    X'&=& a X - b X Y  \label{ModelX} \\
    Y' &=& d X Y - f Y- k X + u + P(t) \label{ModelY},
\end{eqnarray}
where the tumor cells are supposed to be in exponential growth
(which is, however, a good approximation only for the initial
phases of the growth) and the presence of tumor cells implies a
decrease of the "input rate" of lymphocytes. System
(\ref{ModelX})-(\ref{ModelY}) may be rewritten in non-dimensional
form \cite{[Sotolongo]}:
\begin{eqnarray}
    x'&=& \alpha x - x y  \label{ModelAdiX} \\
    y' &=& x y - \frac{1}{\alpha}y - k x + \sigma + p(t) \label{ModelAdiY}
\end{eqnarray}
(in short notation $(x',y')=C(x,y)$). The function $p(t)\ge 0$ is
assumed periodic with period $T$ and it models the effect of immunotherapy. The model has been studied in depth both in the case
of absence of therapy and in the case of therapy by using the test
function $p(t) = 0.5 F(1+cos(4 \pi \nu t))$.

The model shows two equilibria (one of which is tumor-free) and
also unbounded growth. However, the system
(\ref{ModelAdiX})-(\ref{ModelAdiY}) allows negative solutions for
non small $x$, which is not physically acceptable. In fact:
\begin{equation} C(x,0)= (a x, \sigma + p(t) - k x)
\end{equation}
implies that for $x > (\sigma + p_{max})/k$ it is
$C(x,0).(0,-1)>0$, and  $y(t)$ becomes negative in finite times.
Furthermore, the second equilibrium point is a consequence of the
negativity of $\sigma -k x$.

The model in \cite{[Sotolongo]}, though it has this problem of lack
of physical consistency, is, however, of great interest because
the killing of lymphocytes is seen as function of the $x$ variable
Alternatively, the influx of lymphocytes may be thought of as a
function of the entity of the disease, which we will denote as
$Q(x)$. Indeed, it has been observed that in some cases cancer progression may cause generalized immunosuppression. See \cite{[Finn]} and references therein. See \cite{[Finn]} and references therein. Thus, in \cite{[Sotolongo]} it is $
Q(x) = \sigma (1 - (k/\sigma) x) $, which may be read as a first
order Taylor approximation of a more general non-increasing
function.

However a general influx function is only one of the possible
modifications of model (\ref{ModelAdiX})-(\ref{ModelAdiY}): there may be others, which are also biologically reasonable. One might take into the account many factors:
different functional forms for the interaction term, saturation in
the predation term and, mainly, non exponential growth of the
cancer: logistic, gompertzian, generalized logistic etc $\dots$
All these modifications are reasonable and useful. Thus, I think
that it might be useful to define and study  the following general
family of models:
\begin{eqnarray}
    x'&=& x (\alpha f(x) - \phi(x) y)  \label{GenFamilyXo} \\
    y' &=& \beta(x) y - \mu(x) y + \sigma q(x) + \theta(t) \label{GenFamilyYo}
\end{eqnarray}
where:

\begin{itemize}
	\item $x$ and $y$ are the non-dimensionalized numbers of, respectively, tumor cells and of effectors cells of immune system;
	\item $0<f(0) \le +\infty$, $f'(x) \le 0$ and in some relevant
cases we shall suppose that it exists an $0 < \overline{x} \le
+\infty$ such that $f(\overline{x})=0$), $Lim_{x\rightarrow 0^+}{x
f(x)} = 0$. Thus, $f(x)$ summarizes many widely used models of tumor
growth rates, such as the Exponential model: $f(x)=1$ \cite{[W]},
the Gompertz: : $f(x)=Log(A/x)$ \cite{[W],[Marusic]} and its
generalizations \cite{[W],[Marusic]}, the Logistic model:
$f(x)=1-x/A$ \cite{[Marusic]}, the Hart-Schochat-Agur:
$f(x)=x^{-\gamma} \textrm{, } 0<\gamma<1$ \cite{[Zvia]}, the von
Bertanlaffy: $f(x)=x^{-1/3}-b$ \cite{[Marusic],[Vaidya]}, the
Guiot's et al. model: $f(x)=x^{3/4}-b$ \cite{[Guiot]}, the linear
growth model by Bru and coworkers\cite{[Bru]} which may be
written as follows: $f(x)=x^{-1/3}$ (note that it may be
considered a particular case of the von Bertalaffy model and of
the Hart-Shochat-Agur model) etc$\dots$
	\item $\phi(x) >0$, $\phi(0)=1$, $\phi'(x) \le 0$ and $x\phi(x) \rightarrow l \le +\infty$;
	\item $q(x)$ is such that $q(0)=1$ (as a consequence $\sigma = Q(0)$) and it may be non-increasing
or also initially increasing and then decreasing, i.e. we may
assume that either the growth of tumor decreases the influx of
immune cells or that, on the contrary, it initially stimulates the
influx);
	\item $\beta(x) \ge 0$, $\beta(0)=0$ and $\beta'(x) \ge 0$;
	\item $\mu(x)>0$ and $\mu'(x)>0$.
\end{itemize}
For the sake of simplicity we define the following function $\Psi(x)=\mu(x)-\beta(x)$ and write:
\begin{eqnarray}
    x'&=& x (\alpha f(x) - \phi(x) y)  \label{GenFamilyX} \\
    y' &=& - \Psi(x) y + \sigma q(x) + \theta(t). \label{GenFamilyY}
\end{eqnarray}
$\Psi(x)$ is assumed to be positive, otherwise it may be positive in $[0,x_1)\cup(x_2,+\infty)$ with $\Psi(x_1)=\Psi(x_2)=0$. We may assume that it has an absolute minimum in $[0,+\infty)$. We may use $\Psi(x)$ to classify the tumors depending on their degree of aggressiveness against the immune system:
\begin{itemize}
    \item $\Psi(x) >0$: in such a case the ability of destroying immune cells is never won by
    the stimulatory effect on the immune system, therefore the tumor may be indicated as "highly aggressive"/"lowly immunogenic";
    \item Variable sign $\Psi(x)$: since in such a case the destruction of cells may be
    compensated by the stimulatory effect, we will refer to such a tumors as "lowly aggressive"/"highly immunogenic".
\end{itemize}

The above model includes as particular cases the models
\cite{[Stepanova],[Kuznetsov],[Galach],[deVladar]}. For instance, the
Stepanova model \cite{[Stepanova]} is such that $f(x)=1$,
$\phi(x)=1$, $\beta(x)=\beta_1 x$, $q(x)=1$ and
$\mu(x)=\mu_o+\mu_2 x^2$; the de Vladar-Gonzalez model
\cite{[deVladar]} is similar, but: $f(x)= Log(K/x)$.

Note that Nani and Freedman proposed an interesting model of
adoptive cellular immunotherapy in which generic functions are
used \cite{[NaniMbs]}. However, their approach differs from ours
since in their model the proliferation of cells of the immune
systems is not stimulated by cancer cells. In other words in the Nani and Freedman
model the interaction tumor cells - immune system is only
destructive for immune cells. Furthermore, in their model the
"loss rates" are proportional (in our notation we might write
$\mu(x)= \mu(0) + const \phi(x)$).

In the absence of treatment, system
(\ref{GenFamilyX})-(\ref{GenFamilyY}) admits the existence of a
cancer free equilibrium $CF = (0,\sigma / \Psi(0))$.

If $f(0)<+\infty$, we have that if $\sigma > \sigma_{cr}=\alpha \Psi(0) f(0) / \phi(0)$ $CF$ is locally asymptotically stable (LAS), unstable if $\sigma < \sigma_{cr}$.
Biologically, $\sigma > \sigma_{cr}$ means that the immune system works very well and that it is able to destroy small tumors. On the contrary $\sigma \approx 0$ means that there is immunodepression.

Furthermore, when $\phi(x)= constant = \varphi$ and $\Psi(x) \le
\Psi^* < +\infty  $, if $\sigma > \sigma^* = \alpha f(0) \Psi^* /
(q_{min} \varphi)$ it follows that $CF$ is globally asymptotically
stable ($GAS$). In fact, from $y'=-\Psi(x)y + \sigma q(x) \ge
-\Psi^* y +\sigma q_{min}  $ if follows that asymptotically $y(t)
\ge \sigma q_{min}/\Psi^*$. As a consequence, asymptotically $x'
\le (\alpha f(0) -\varphi (\sigma q_{min} / \Psi^*)  )x $, i.e. if
$\sigma > \sigma^*$  it is $x(t) \rightarrow 0  \Rightarrow y(t)
\rightarrow \sigma / \Psi(0)$.

A relevant problem, up to now, is that the immunotherapeutic agents are characterized
by strong toxicity, thus $\sigma > \sigma^*$ might be too biologically high, even in
cases in which when it is mathematically small.

If $f(0)=+\infty$, as in the Gompertzian case (used, for example,
in \cite{[deVladar]}) and in other tumor growth models, then $CF$ is
unstable anyway (as previously stressed for the particular model
\cite{[deVladar]}) because in such a cases the derivative of $x
f(x)$ at $x=0$ is $+\infty$. In the light of \cite{[deVladar]} and
of our generalization, this implies that the immune system would
never be able to totally suppress even the smallest tumor cell
aggregates, which is a very strong inference. This instability
result deserves some comments because it has deep medical
implications: the impossibility to completely recover from
any type of tumors whatsoever. On the contrary, it is commonly held that the
immune system may be able, in some cases, to kill a relatively small
aggregate of cancer cells. In the background of all cancer
therapies (which are of finite duration) there is the implicit
hypothesis that the drug will kill the vast majority of the malignant
cells and that the relatively few residual cells may in some
cases be killed by the immune system \cite{[Bonadonna]}. Accepting
this hypothesis, the equilibrium $CF$ should have the possibility
to be LAS and, as a consequence, for small $x$ the function $f(x)$
should be bounded.

The modeling of cancer by means of the Gompertz law of
growth was introduced in early sixties by A. K. Laird
\cite{[Laird1],[Laird2]}. She conducted pioneering data-fitting work using a vast amount of real data and justified the law in terms of increasing
mean generation time. There is much research showing that the
Gompertzian model fits data well from experimental and in vivo
tumors
\cite{[Olea],[Bassukas],[Parfitt],[Rygaard],[Ballangrud],[Cameron],[Castro]}.
From a theoretical point of view,  Gyllenberg and Webb
\cite{[GW]}, Calderon and Kwembe \cite{[Calderon]}, Calderon and
Afenya \cite{[Afenya1],[Afenya2]} proposed physico-mathematical
justification of the Gompertz model. Furthermore, some
interesting physical properties of the Gompertz model have been
elucidated by Konarski and Molski \cite{[Molski]} and by Konarski
and Waliszewski \cite{[CSF]}.

However, the doubling time of a population of cells cannot be
lower than the minimal time needed by a cell to divide, which is
obviously non-null. This biological constraint is in contrast with
the unboundedness of $f(x)$ in the Gompertz and other models, as stressed by Wheldon \cite{[W]}.
More recently, inconsistency at low number of cells have been
recognized by Castorina and Zappala' in their derivation of the
Gompertizan model based on methods of statistical mechanics
\cite{[CZ1],[CZ2]}. They showed that the validity of the
Gompertz model starts above a minimum threshold for the number
of cells, whereas under the threshold there is exponential growth.
In other words, they derived biophysically the Gomp-Ex model
proposed on biological ground in \cite{[St],[W]}. Using data from multicellular tumor spheroids, Marusic and coworkers performed a systematic comparison of many models \cite{[Marusic]}, which showed that Gompertz's model fitted their data very well, but slightly less well than the Piantadosi model \cite{[Piantadosi]}, which has finite $f(0)$. Furthermore, in their fittings, it was not possible to discriminate between the pure Gompertz model and the Gomp-Ex
model. Demicheli and coworkers used Gomp-Ex model on in vitro and in vivo data obtaining results strongly supporting this model \cite{[Dem]}. Other comparisons may be found in \cite{[Afenya1],[Vaidya]}. Moreover, in general, van Leeuwen and Zonneveld \cite{[Ingeborg]} claims that it may be not possible
to discriminate between exponential, logistic and gompertzian models
in the early phases of growth. Recent experimental studies conducted by Bru
and coworkers support an initial phase of
exponential growth \cite{[Bru]}. Summarizing, I consider the results
by de Vladar and Gonzalez (and our extensions) to be very valuable,
but they may be read in a dichotomic way:
\begin{itemize}
    \item A tumor is permanent: the innate immune surveillance is never able to completely eradicate even the smallest tumor.
    \item Since there is relevant evidence that the immune system is able in some cases  to eliminate small tumors \cite{[Dunn2002],[Dunn2004]} (as we will see in following sections, the ability of eradicate the disease or not depends on initial conditions), the properties of the de Vladar-Gonzalez model (and of our extension) may be seen as an evidence that Gompertzian and other models characterized by $f(0)=+\infty$ are not appropriate for very small tumors, in coherence with \cite{[W],[CZ1],[CZ2],[Bru]}.
\end{itemize}

In case of the absence of influx of immune cells ($q(x)=0$) and for laws of growth in which $\overline{x}$ exists, there is a different particular equilibrium point, which we shall call "immune free": $IF=(\overline{x},0)$, which is LAS.

Other multiple non null equilibria may be found by finding the positive intersection of the two nullclines:
\begin{eqnarray}
    y_C(x)&=& \alpha \frac{ f(x)}{\phi(x)}  \label{ncC} \\
    y_I(x) &=& \frac{\sigma q(x)}{\Psi(x)}. \label{ncI}
\end{eqnarray}
The functions $ y_C(x)$ and $ y_I(x)$ are useful in the determination of the LAS of the equilibria , since the characteristic polynomial of the Jacobian, calculated at a given equilibrium point $(x_e,y_e)$, is:
\begin{equation}\label{charpol}
\lambda^2 +\Big(\Psi(x_e)-x_e\phi(x_e)y_C'(x_e)\Big)\lambda + \Psi(x_e) x_e \phi(x_e) \Big(-y_C'(x_e)+y_I'(x_e)\Big) =0.
\end{equation}
So the LAS condition is:
\begin{equation}\label{lascond}
y_C'(x_e) < \frac{\Psi(x_e)}{x_e\phi(x_e)} \textrm{   AND }  y_I'(x_e)> y_C'(x_e).
\end{equation}
Note that the first part of the AND condition is automatically fulfilled when $y'_C(x) \le 0$ (because $x_e$ cannot lie in an interval where $\Psi(x)<0$), whereas the second part has a straightforward geometrical interpretation.

Finally, it is interesting to note that the above family of model may admit limit cycles if $f(x)=1$ (exponential growth) and $q(x)$ is identically null for $x>x_q$ with $x_q<x_1$. In fact, in such a case there is the equilibrium point $(x_1,\alpha)$ whose characteristic polynomial is:
\begin{equation}\label{charpollc}
\lambda^2 + h^2 = 0 \textrm{, } h^2 :=  -x_1 \Psi'(x_1)\alpha > 0
\end{equation}
In effect, some cases of sustained oscillations (or slow oscillations with very small damping) have been reported in the medical literature \cite{[Kennedy],[Vodopick],[Tsao]}. Periodic solutions in absence of influx of immunocompetent cells are predicted also in \cite{[Kirschner]}.

On the contrary, if $y'_C(x) \le 0$ (for example when $\phi(x)$ is constant), by applying the Dulac-Bendixon theorem with multiplicative factor $1/(x y \phi(x))$ (as in the specific models \cite{[Kuznetsov],[Galach]}) one obtains that the presence of limit cycles is not possible. In fact:
\begin{equation}\label{dulacth} Div\Big( \frac{1}{x y \phi(x)}(x'(x,y),y'(x,y))\Big) =  \frac{\alpha y_C'(x) }{y} - \sigma \frac{q(x)}{x \phi(x) y^2} < 0 \end{equation}
\subsection{The global behavior}\label{global}
In some important cases, it is possible to study the global
behavior of the family, by means of differential inequalities and
of the Poincare-Bendixon trichotomy \cite{[Thieme]}. We may state
the following simple propositions:

\begin{enumerate}
    \item {\em When $\Psi(x)>0$ and  $f(x)=1$ and $y_C'(x) \ge 0$, if it is $y_I(x)<y_C(x)$ then
    $x(t)\rightarrow +\infty$}.  {\bf Proof: }Let us define
    $y_I^{MAX}:= Max_{x \in \R_+} y_I(x) $ and $x_M$ such that $y_I(x_M)=y_I^{MAX}$. If it is $ y_I(x)<y_C(x) $ it is easy to show that the set
    $H =\Big\{(x,y) | x>x_M \textrm{ AND } 0 \le y \le y_I^{MAX} \Big\}$ is positively invariant
    and adsorbing. Thus, since in $H$: $x' \ge x \phi(x) (y_C(x_M) - y_I^{MAX}
    )>0$, it follows readily that $  x(t) \rightarrow + \infty$;

    \item {\em It $\Psi(x)>0$, it exists $\bar{x}$ such that  $f(\bar{x})=0$ , $y_C'(x) < 0$  and there is a unique LAS equilibrium point $S=(x_e,y_e)$ , then $S$ is GAS.}
    {\bf Proof: } Let us define $y_I^{MAX}:= Max_{x \in [0,\overline{x}]} y_I(x) $ and
    $y_I^{min}:= Min_{x \in [0,\overline{x}]} y_I(x) $.
    Furthermore, if $f(0)> y_I^{MAX}$ let it be
    $\tilde{x}=y_I^{-1}(y_I^{MAX})$, if $f(0)\le y_I^{MAX}$ let it
    be $\tilde{x}=0$. Since $  \Psi(x) (y_I^{min} -y) \le y' \le \Psi(x)( y_I^{MAX} -y ) $
    it is easy to see that the set $R= \Big\{(x,y) | \tilde{x} < x \le \overline{x} \textrm{ AND } y_I^{min} \le y \le y_I^{MAX} \Big\}$
    is positively invariant and adsorbing and contains $S$. Since we have ruled out the
    possibility that there may be limit cycles, as a consequence  $S$ is $GAS$.

    \item {\em When $\Psi(x)>0$ and  $y_C'(x)$ is non-constant and there is a unique LAS equilibrium
    point $S=(x_e,y_e)$, if it holds also that
    \begin{equation} y_C^{Max} >  y_I^{Max} \end{equation}  then $S$ is $GAS$.}  {\bf Proof: }
    When $f(x)$ is unbounded, one may see that,   there may be a relative minimum followed by a
    relative maximum in $(0, \overline{x})$. On the contrary, when $f(x)$ is bounded, there
    is an absolute maximum.  Calling now $x^*$ the point in which $y_C(x)$ is (absolutely or
    relatively) maximum,  one has that
    $R^*=\Big\{(x,y) | x^* \le x \le \overline{x} \textrm{ AND } y_I^{min} \le y \le y_I^{MAX} \Big\}$
    is positively invariant and adsorbing,  contains $S$. Since in $R^*$  it is
    $y_C^{'}(x) \le 0$ ( which implies that closed orbits are ruled out), as a consequence,
    $S$ must be GAS.

    \item {\em When $\Psi(x)>0$ and $y_I(x)>y_C(x)$ for $x \in [0,\bar{x}]$ then $CF$ is GAS.}
    {\bf Proof: } It is a particular case of proposition 2.

    \item {\em If $\Psi(x)>0$, there does not exist a $\bar{x}$ such that  $f(\bar{x})=0$ , $y_C'(x) < 0$  and there is a unique LAS equilibrium point $S=(x_e,y_e)$ , then $S$ is GAS.}
    {\bf Proof: } Let us define $y_I^{MAX}:= Max_{x \in [0,+\infty)} y_I(x) $.
    Let us consider a point $P_o=(x_o,0)$ with $x_o>x_e$, and the orbit starting
    from it, which intersects the curve $y_C(x)$ in the point
    $P_a=(x_a,y_C(x_a))$. Let us consider the following points
    $P_b=(x_a,y_I^{MAX})$, $P_c=(0,y_I^{MAX})$ and $P_d=(0,0)$.
    The arc of orbit $\widehat{P_oP_a}$ and the straight segments
    $\overline{P_aP_b}$,$\overline{P_bP_c}$,$\overline{P_cP_d}$
    and $\overline{P_dP_o}$ bounds an invariant set for our
    system. As a consequence of the Bendixon-Poincare' tricothomy we have
    that S is GAS.

    \item {\em When $\Psi(x)$ has variable sign, and $f(x)$ is bounded and $y_I(x) > y_C(x) $ then $CF$ is $GAS$.} {\bf Proof: } the set $X = \Big\{(x,y) | 0 < x \le \overline{x} \textrm{ AND } y \ge 0 \Big\}$ is positively invariant and adsorbing and in it closed orbits are impossible, as we have seen. However it is not a bounded set, so we have to show that all the orbits starting in $X$ are bounded. Firstly, we notice that it cannot be  $y(t) \rightarrow +\infty$, since in such a case, being $ x' = x (\alpha f(x) - y(t) )$, it would be $ x(t) \rightarrow 0 \Rightarrow y(t) \rightarrow \sigma / \Psi(0) $. Furthermore hypothetical solutions such that $ minlim_{ t \rightarrow +\infty} \textrm{ } y(t) =0 $ and $Maxlim_{t \rightarrow +\infty} \textrm{ }  y(t) = +\infty $ are not possible since the set $A= \Big\{(x,y) | 0 < x \le x_1 \textrm{ AND } y \ge y_c(x) \Big\}$ is positively invariant. As a consequence of these properties, thanks to the Bendixon-Poincare' trichotomy, $CF$ is GAS.

    \item {\em When $\Psi(x)$ has variable sign, there is $\bar{x}$ such that $f(\bar{x})=0$,
    $y_C(x)<0$ and there is a unique LAS equilibrium point $S$ then $S$ is $GAS$.}
    {\bf Proof: } the set $X = \Big\{(x,y) | 0 < x \le \overline{x} \textrm{ AND } y \ge 0 \Big\}$
    is positively invariant and adsorbing and in it closed orbits are impossible, as we have
    seen. However it is not a bounded set. Let us consider $y_I(x)$: it is such that it is
    split in two branches: $y_I^{right}(x)$ for $x_2 \le x \le +\infty$ (which has no intersections
    with $y_C(x)$) and $y_I^{left}(x)$ for $0 \le x < x_1$ (on which $S$ lies). Let us consider
    a point $P_i=(x_i,y_i)$ lying on  the curve $(x,y_I^{right}(x))$ and having $y_i > y_C(0) > y_C(x_2)$. Let the orbit starting from $P_i$ intersect the graph $(x,y_I^{left}(x))$ in a  point $P_f=(x_f,y_f)=(x_f,y_I^{left}(x_f))$ (note that it is $y_f>y_i$). Let us define the following points: $P_A=(0,y_f)$, $P_B=(0,\bar{x})$ and $P_C=(\bar{x},y_i) $. It is easy to see that segment of orbit $\widehat{P_i P_f}$ and the straight segments $\overline{P_f P_A}$,$\overline{P_A O}$, $\overline{O P_B}$, $\overline{P_B P_C}$ and $\overline{P_C P_i}$ bound an invariant set $\Omega$ for our dynamical system. As a consequence, thanks to the Bendixon-Poincare trichotomy, $S$ is GAS.

     \item {\em When the sign of $\Psi(x)$ is variable, there is no $\bar{x}$ such that $f(\bar{x})=0$,
    $y_C(x)<0$ and there is a unique LAS equilibrium point $S$ then $S$ is $GAS$.} {\bf Proof: } The         proof is easily obtained by applying methods of propositions 7 and 5 to find a bounded positively invariant set surrounding S.
    
\item {\em When $\Psi(x)>0$ and $q(x)=0$ then $\forall (x(0),y(0))$ it is $y(t) \rightarrow 0^+$. Furthermore, in accordance with the growth law $f(x)$, either the tumor tends to an equilibrium value or it grows unbounded}. {\bf Proof: } Let us define $\Psi_{min}=min_{x \in \R_+}\Psi(x)$. If $q(x)=0$ it is $y'=-\Psi(x)y \le \Psi_{min} $ $ \Rightarrow y(t) \rightarrow 0^+$. Thus, the equation for $x(t)$ becomes asymptotically autonomous, so that, depending on $f(x)$, either $x(t) \rightarrow +\infty$ or  $x(t) \rightarrow \bar{x}$  (i.e. in this case the equilibrium $IF=(\bar{x},0)$ is GAS).
    
    \item {\em When $\Psi(x)>0$ and  $f(x)=1$ and $\phi(x)= const =\varphi$, and there are two equilibria $S=(x_e,y_e)$ (LAS) and $U=(x_u,y_u)$ (unstable) and there is a separatrix curve $y=\Sigma(x)$ which does not join S to U, then there are two sets $A$ and $B$ such that if $(x(0),y(0)) \in A$ then $(x(t),y(t))\rightarrow S$, whereas if $(x(0),y(0)) \in B$ then  $x(t)\rightarrow +\infty$}. {\bf Proof: } Let us define
    $y_I^{MAX}:= Max_{x \in \R_+} y_I(x) $ and $x_{\Sigma}=\Sigma^{-1}(y_I^{MAX})$ . As a consequence, the set $A=\Big\{(x,y) | 0< x < x_{\Sigma} \textrm{ AND } Min(0,\Sigma(x)) \le y \le y_I^{MAX} \Big\}$ is positively invariant and in it there are no closed orbits, so if $(x(0),y(0)) \in A$ then $(x(t),y(t))\rightarrow S$. It is easy to show that given a $ y_I(x_{\Sigma})/\varphi<\varrho < \alpha /\varphi $ also the set
    $B =\Big\{(x,y) | x>x_{\Sigma} \textrm{ AND } 0 \le y \le \varrho \Big\}$ is positively invariant. Thus, since in $B$: $x' \ge x (\alpha - \varrho)$, it easily follows that $  x(t) \rightarrow + \infty$;

    \item {\em Let it be $\Psi(x)>0$, $y_c'(x) \le 0$ and it exists $\bar{x}$ such that $y_C(\bar{x})=0$ Let there be 4 equilibria CF (unstable), $S_l=(x_e,y_e)$ (LAS), $U=(x_u,y_u)$ (unstable) and $S_r=(x_e,y_e)$ (LAS), and let there be a separatrix curve $y=\Sigma(x)$ which does not join $S_l$ or $S_r$ to U, then there are two sets $A$ and $B$ such that if $(x(0),y(0)) \in A$ then $(x(t),y(t))\rightarrow S_l$, whereas if $(x(0),y(0)) \in B$ then  $(x(t),y(t))\rightarrow S_r$}.  {\bf Proof: } 
    As in the previous proposition $A=\Big\{(x,y) | 0< x < x_{\Sigma} \textrm{ AND } Min(0,\Sigma(x)) \le y \le y_I^{MAX} \Big\}$ is positively invariant and in it there are no closed orbits, so if $(x(0),y(0)) \in A$ then $(x(t),y(t))\rightarrow S_l$. In this case $B=\Big\{(x,y) | 0< x < \bar{x} \textrm{ AND }  0 \le y \le  Min(\Sigma(x),y_I^{MAX})  \Big\}$, and it is positively invariant as well, and with no closed orbits in it. As a consequence: if $(x(0),y(0)) \in B$ then  $(x(t),y(t))\rightarrow S_r$ ;

\end{enumerate}

{\bf Remark} {\em A consequence of the fourth proposition is that
if $y_I'(0)>0$ (or $ y_I'(0)=0 $ AND $y_I''(0)>0$) then $ \sigma >
\sigma_{cr}$ is a sufficient condition for the GAS of the $CF$
equilibrium. }

In case of multiple equilibria with $\phi(x) = const$ it may be useful to transform (\ref{GenFamilyX})-(\ref{GenFamilyY}) to a nonlinear oscillator. In fact by setting $z= Log(x)$ it is easy to see that the original family becomes:
\begin{equation}\label{osc}
z'' + (\bar{\psi}(z)-\bar{f}'(z))z' + \varphi\sigma \bar{q}(z)- \bar{\psi}(z)\bar{f}(z) = 0
\end{equation}
where $\bar{\psi}(z)=\psi(E^z)$ etc$\dots$
By defining the damping coefficient:
\begin{equation}\label{damp}
2 \nu(z) = (\bar{\psi}(z)-\bar{f}'(z))
\end{equation}
and the pseudo-potential:
\begin{equation}\label{pot}
U(z) = \int_{0}^{z}{\Big(\varphi\sigma \bar{q}(s)- \bar{\psi}(s)\bar{f}(s)\Big)ds}
\end{equation}
and the total pseudo-energy:
\begin{equation}\label{tot}
E_{tot} = \frac{(z')^2}{2} + U(z)
\end{equation}
it follows immediately that when $\nu(z)>0$:
\begin{itemize}
    \item Let it be $\bar{x} < +\infty$ and let there be three equilibria $z_l < z_c < z_r$ which are, respectively LAS, unstable and again LAS. Let it be $E_{tot}(0)<U(z_c)$, then $z(0)<z_c$ $\Rightarrow$ $z(t) \rightarrow z_l$, whereas  $z(0)>z_c$ $\Rightarrow$ $z(t) \rightarrow z_r$;
    \item Let it be $\bar{x} = +\infty$ and let there be two equilibria $z_l < z_c $ which are, respectively LAS and unstable. Let it be $E_{tot}(0)<U(z_c)$, then $z(0)<z_c$ $\Rightarrow$ $z(t) \rightarrow Z_l$, whereas  $z(0)>z_c$ $\Rightarrow$ $z(t) \rightarrow +\infty$.
\end{itemize}

\section{On immunotherapies}
\subsection{Therapy schedulings}
A realistic anticancer therapy may be modeled with sufficient
approximation as constant (e.g. via a constant intravenous
infusion) or periodic (e.g. the agent is delivered each day as a
bolus):
\begin{equation} \theta(t) = \theta_m + \Omega(t) \ge  0 \textrm{, } \theta(t+T)=\theta(t), \theta_m = \frac{1}{T}\int_0^T{\theta(t)dt}
\end{equation}
For humans, typical periods ranges between $8$ hours to $7$ days
\cite{[Seminar],[VHR]}. A particular case of periodic therapy is
pulsed therapy, i.e. a therapy which induces an instantaneous
increase of the number of lymphocytes:
\begin{equation} \theta(t) = \gamma \sum_{n=0}^{+\infty}{\delta(t-n T)}
\end{equation}
In the case of constant infusion therapy (CIT)
($\theta(t)=\theta_m$) by defining:
\begin{equation}
\widehat{\sigma}:=\sigma  + \theta_m, \widehat{q}(x):=\frac{\sigma  + \theta_m}{\widehat{\sigma}}
\end{equation}

\textbf{Remark}\textit{ In the next subsections some asymptotic
analyses of therapies shall be conducted. The meaning of the underlying $t \rightarrow +\infty$ limits
is the following: the therapies are administered for a time interval $[0,t_f]$ which is finite but
sufficiently high to guarantee that the number of cancer cells is zero or
that other targets have been reached.}

\subsection{Continuous infusion therapy}
All the considerations we have done the absence of
therapy hold also in case of CIT. In particular, for
$f(0)<+\infty$, the condition for the LAS of the cancer-free
equilibrium is:
\begin{equation} \sigma + \theta_m > \sigma_{cr} \end{equation}
Because of the co-presence of other equilibria, the above
criterion is not global, i.e. the immunotherapy is not able to
guarantee the disease eradication from whatever initial values
$(x(0),y(0))$. However, observing that  in models in which
$\Psi(x)>0$:
\begin{equation}
    y_I^{with therapy}(x) = \frac{\sigma q(x)+\theta_m}{\Psi(x)} > y_I^{no therapy}(x)
\end{equation}
(e.g. in Stepanova's model with low $\mu_1$) it happens that,
roughly speaking, the stable equilibrium size of the cancer
becomes smaller and the unstable equilibria greater, so that the
basin of attraction of the unbounded solution is reduced.

\begin{figure}[ht]
 \centering
 \includegraphics[width=7.2cm]{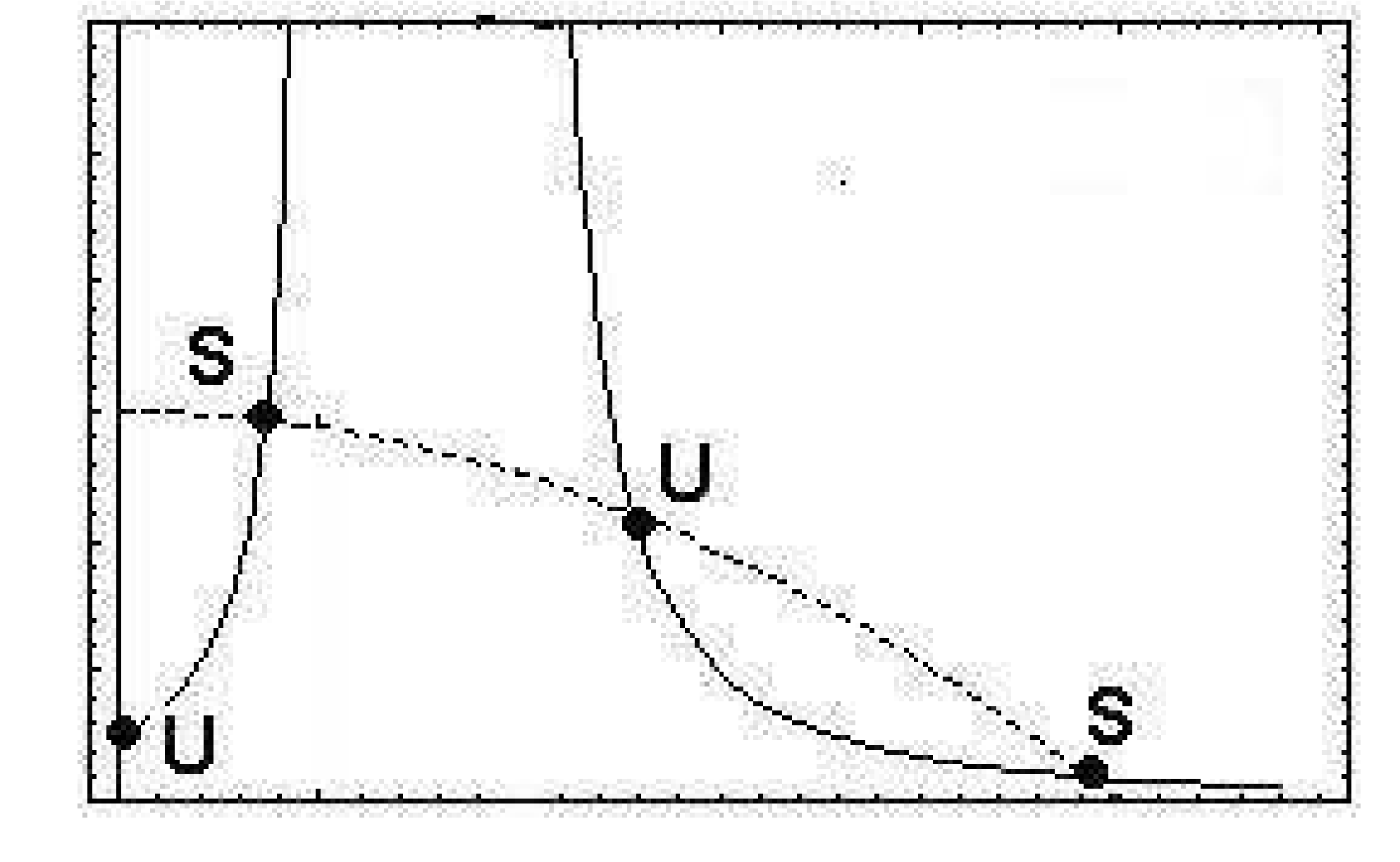}
 \includegraphics[width=7.2cm]{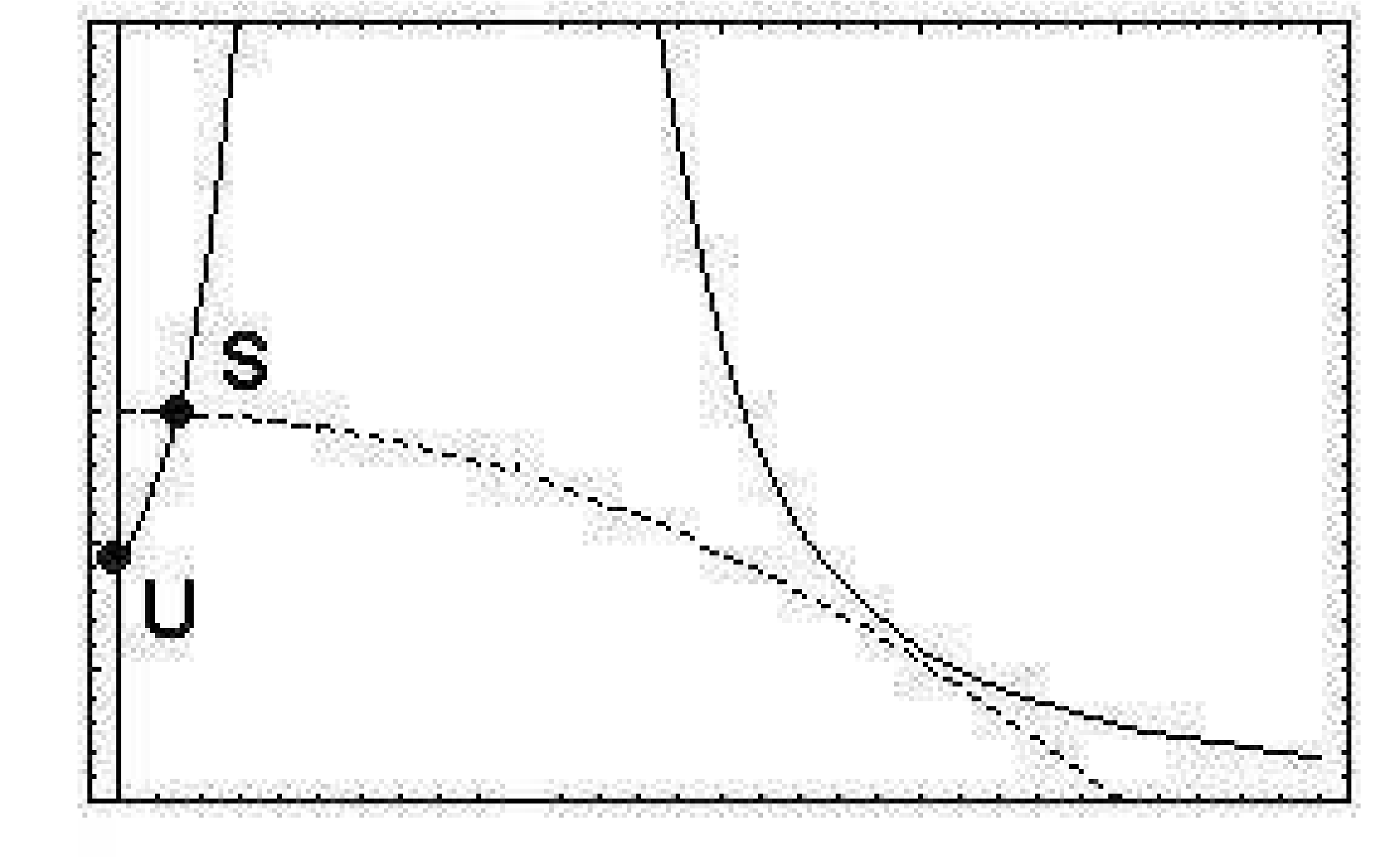}
 \includegraphics[width=7.2cm]{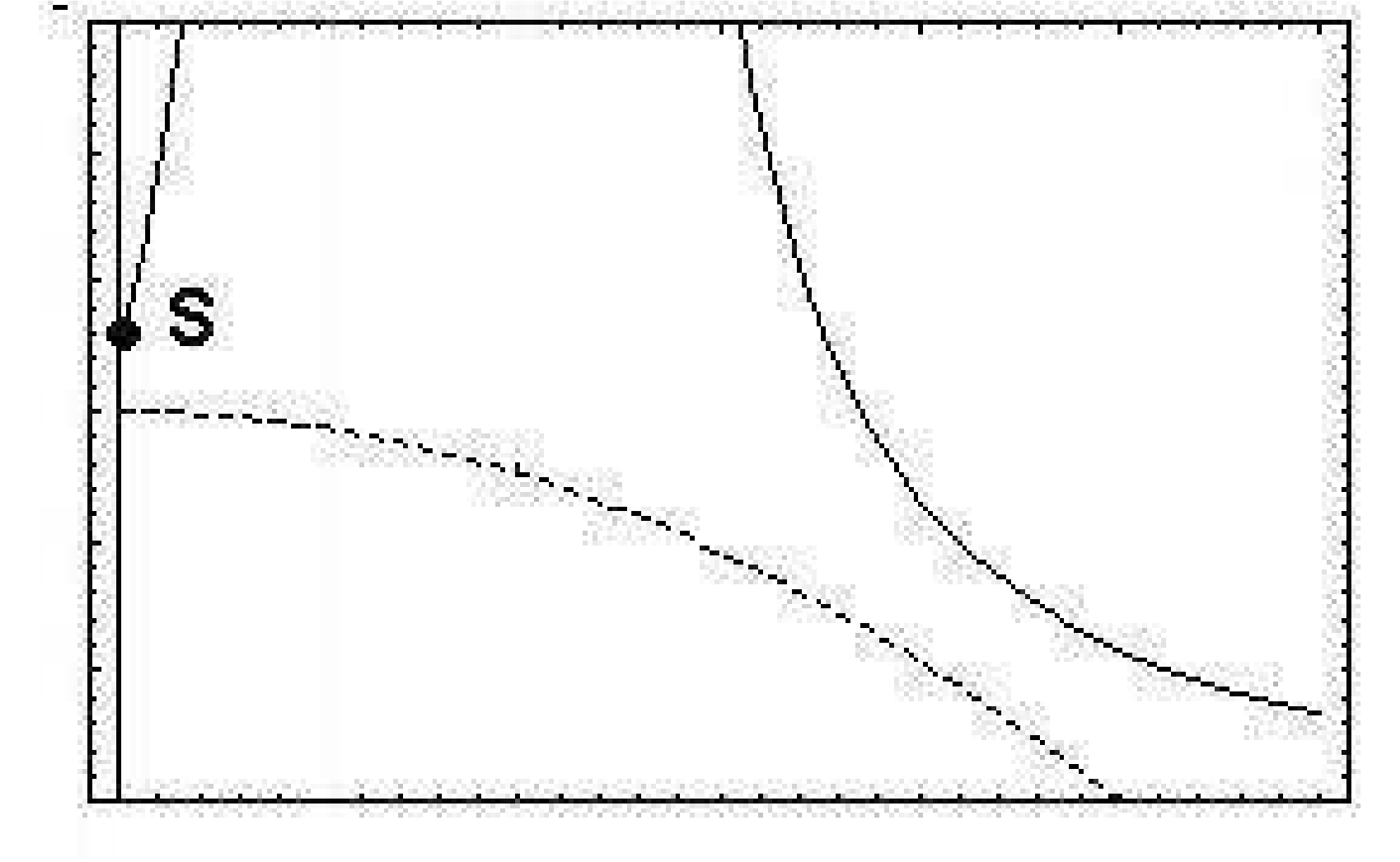}
  \caption{Illustration of the effect of a CIT on a typical configuration in a lowly aggressive
  tumor. The case is shown in which $\theta_r<\theta_{cf}$. $y_I(x)$ is plotted as a solid line, whereas $y_c(x)$ is dashed. The equilibria are plotted as black points and they are labeled $U$ when unstable, otherwise $S$. First subfigure: figure: in the absence of therapy there are four equilibria among which CF. Second subfigure: with a therapy with $\theta_r<\theta<\theta_{cf}$ CF is unstable and coexists with a microscopic tumor equilibrium which is GAS. Fourth subfigure: for a high dose therapy $\theta>\theta_{cf}$ CF becomes GAS.}
  \label{lb}
\end{figure}

\begin{figure}[ht]
 \centering
 \includegraphics[width=7.2cm]{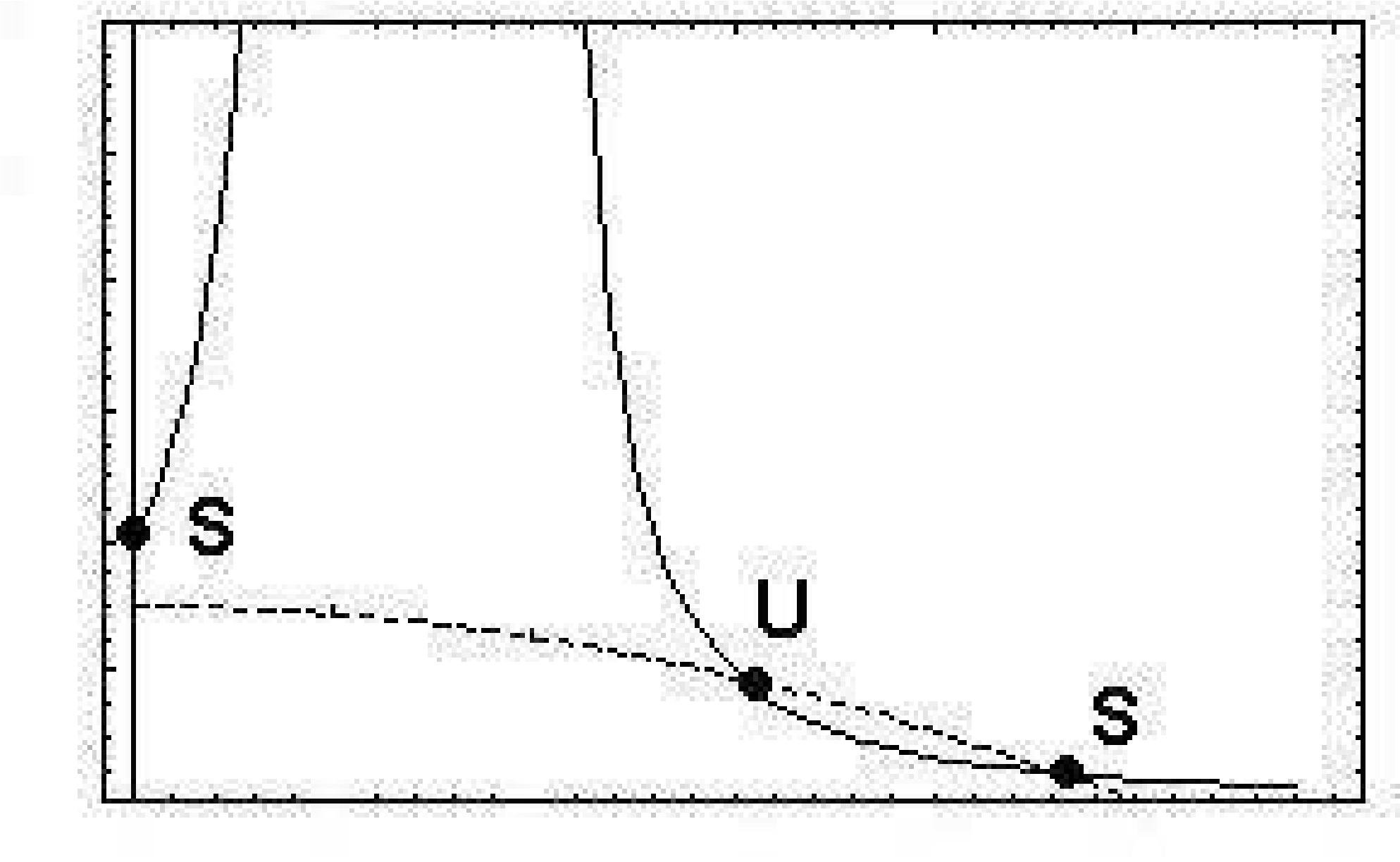}
  \caption{Illustration of the effect of a CIT in a low aggressive tumor for $\theta_{cf}<\theta_r$ and $\theta_{cf}<\theta<\theta_r$. Symbols as in figure \ref{lb}}
  \label{lbpat}
\end{figure}

\begin{figure}[ht]
 \centering
 \includegraphics[width=7.2cm]{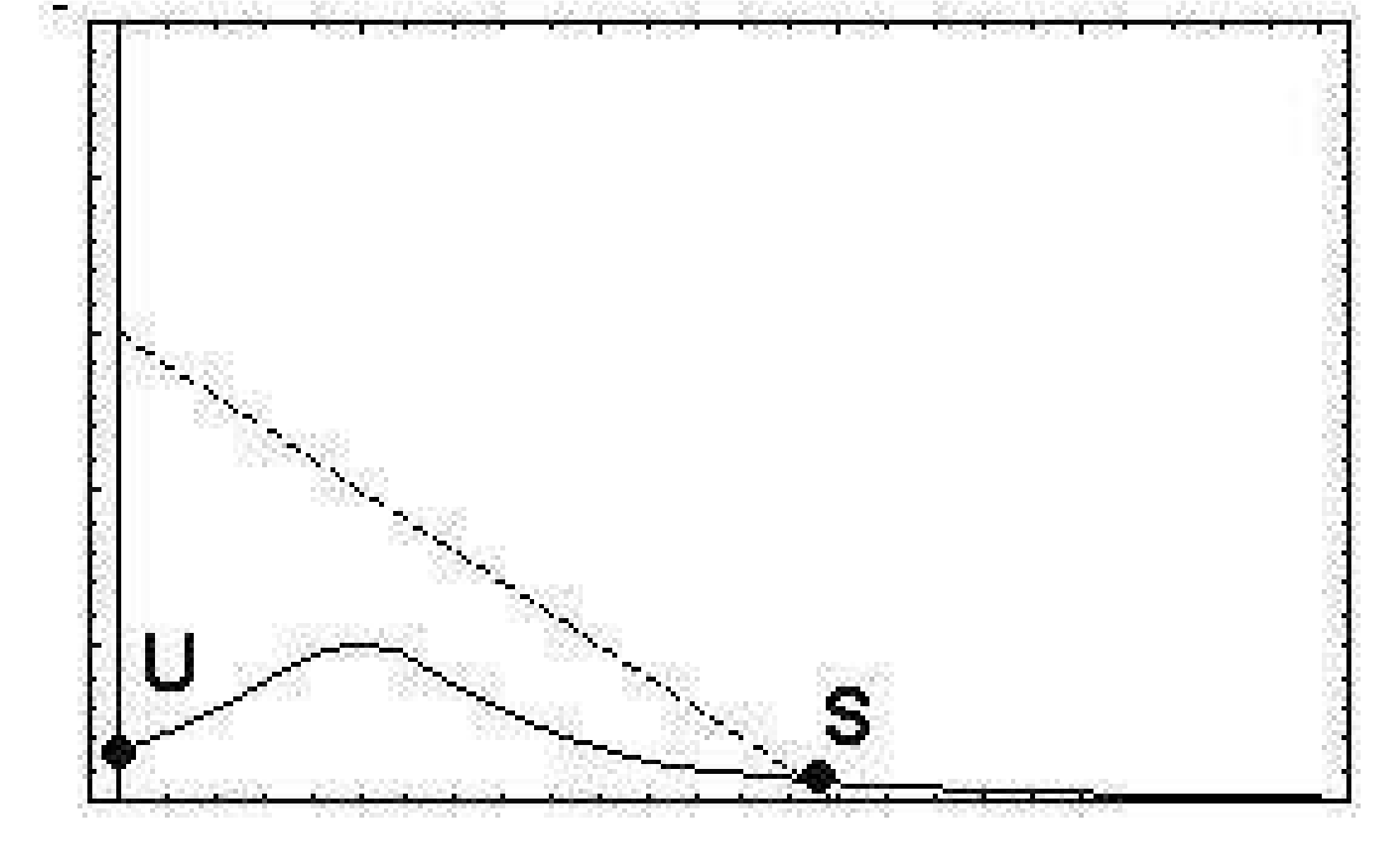}
 \includegraphics[width=7.2cm]{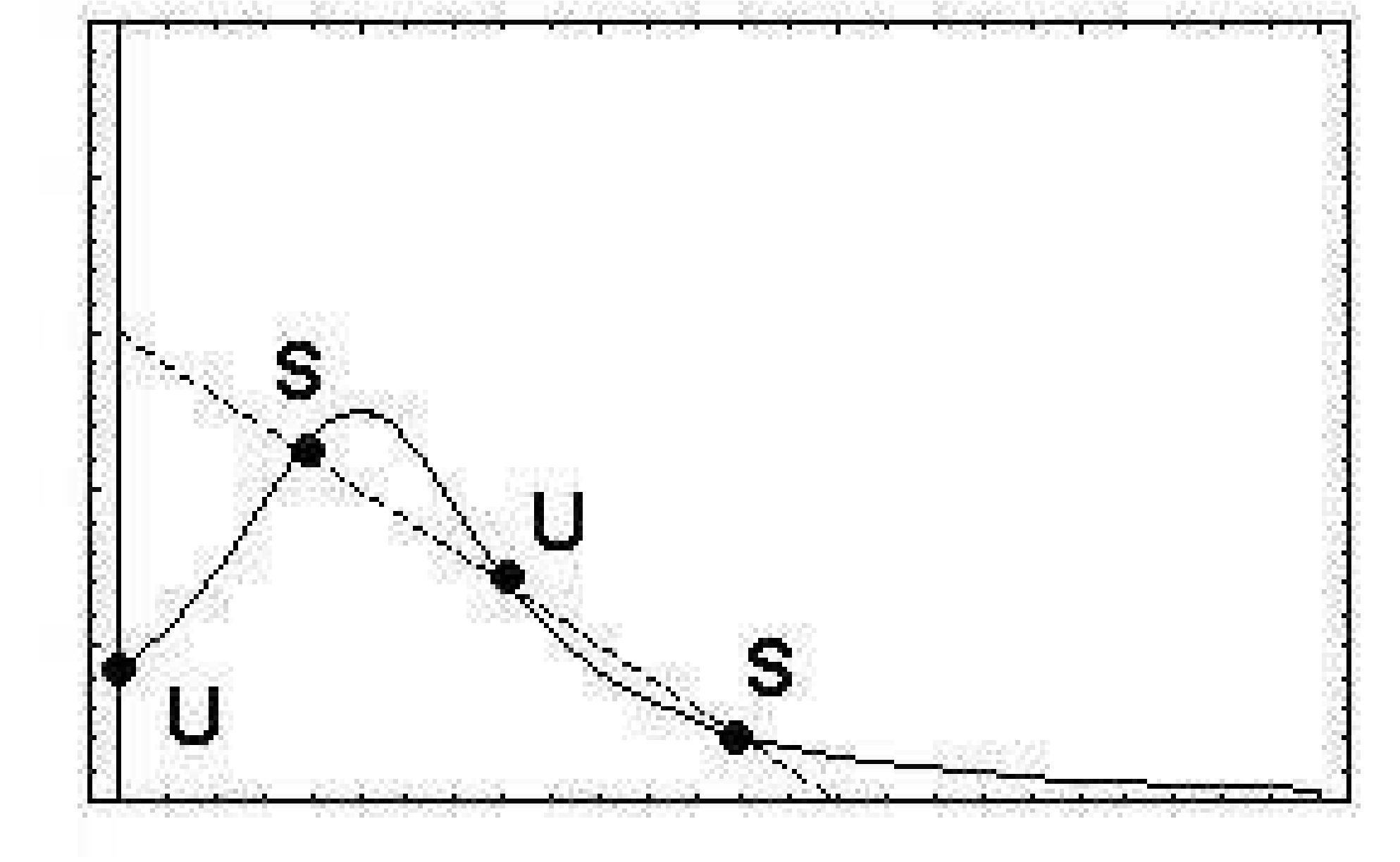}
 \includegraphics[width=7.2cm]{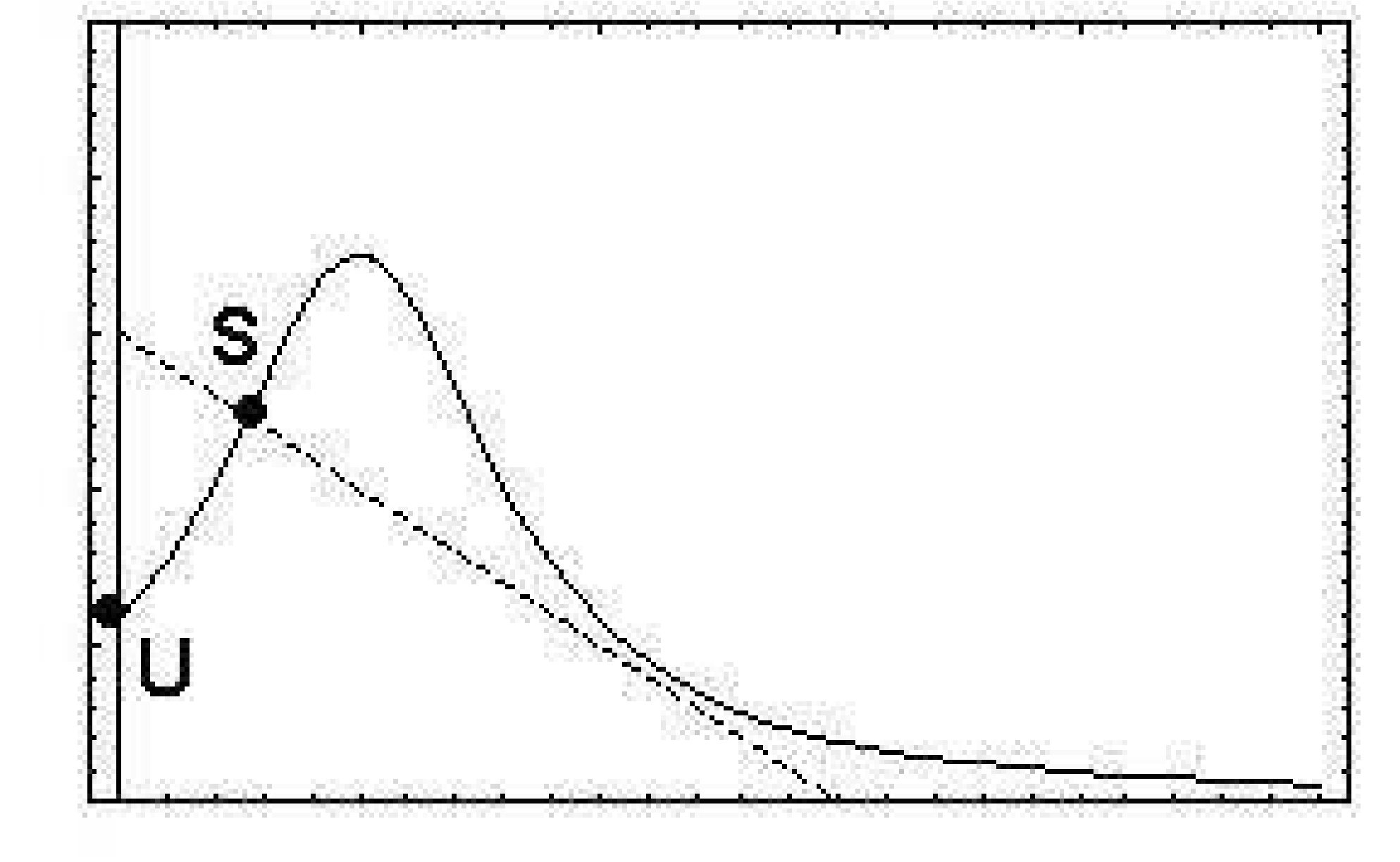}
 \includegraphics[width=7.2cm]{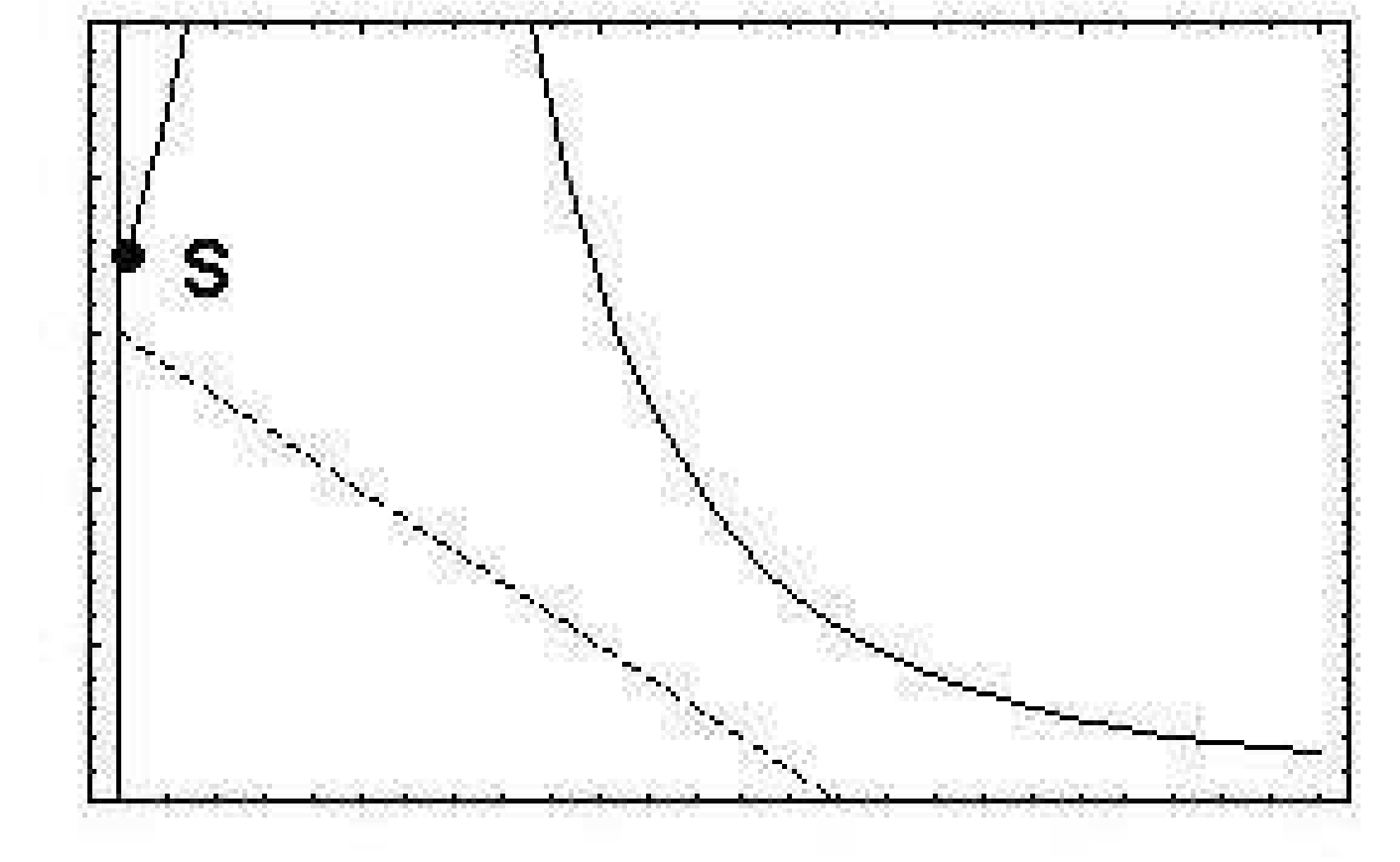}
  \caption{Illustration of the effect of a CIT in an aggressive tumor for increasing values of the CIT.}
  \label{lh}
\end{figure}

\begin{figure}[ht]
 \centering
 \includegraphics[width=7.2cm]{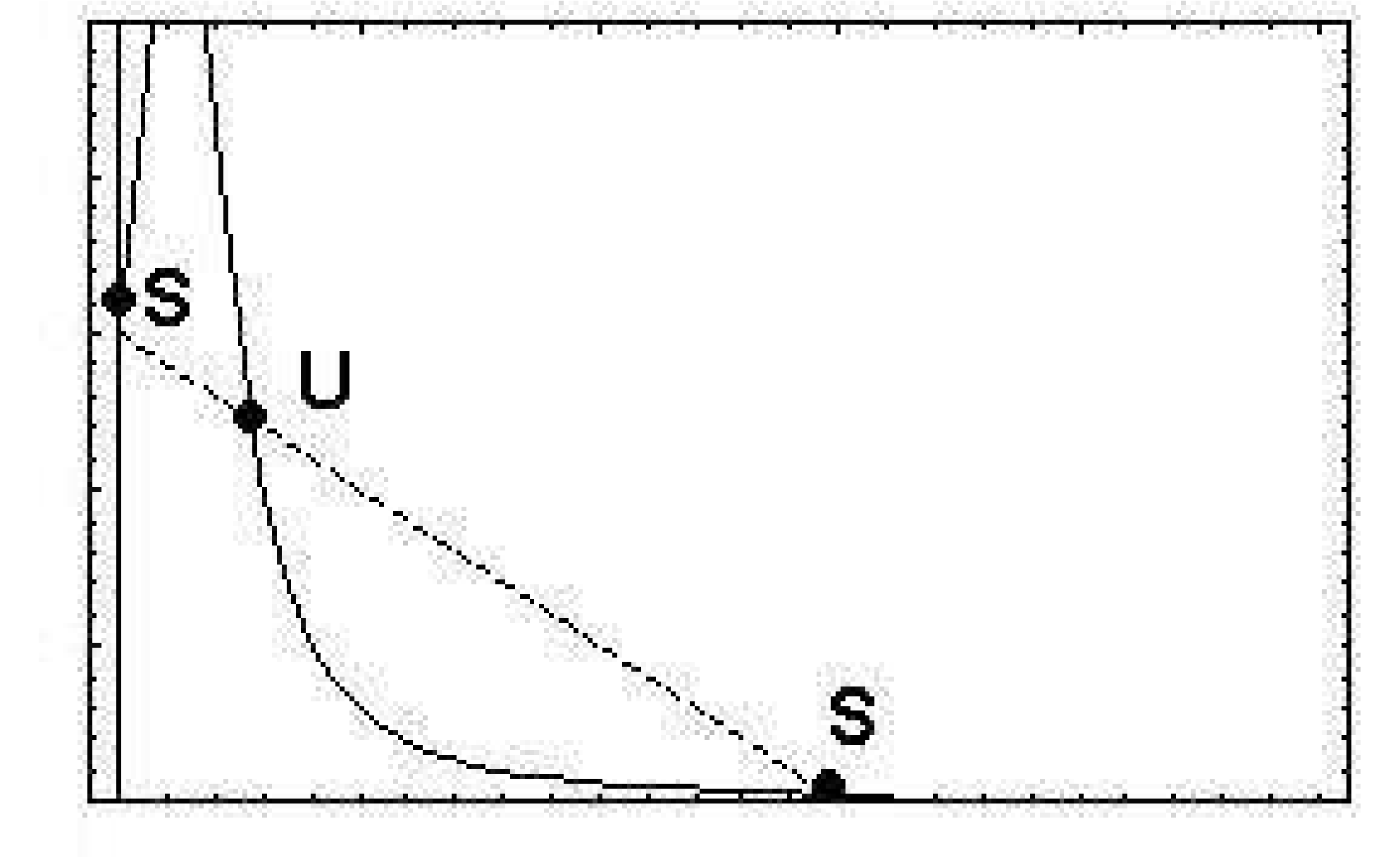}
  \caption{Illustration of the effect of a CIT in an aggressive tumor, similar to figure \ref{lh}, but with LAS CF coexisting with two other equilibria ($\theta_{cf}$ "low").}
  \label{lhpat}
\end{figure}

Let us consider now some typical situations in case of $y_C'(x)<0$:

\begin{itemize}
\item Non aggressive tumor (i.e. $\Psi(x)\le 0$ in $[x_1,x_2]$). In such a case, in absence of
therapy there may be in the most complex case 4 equilibria: CF
(unstable), a small tumor equilibrium $E_{micro}^o$ (LAS), a
macroscopic equilibrium $E_{MACRO}^O$ (LAS) and an intermediate
unstable equilibrium $E_U^O$, as in figure \ref{lb}-subplot 1.
$E_{micro}^o$ is determined by the intersection between $y_C(x)$
and the branch $y_I^l(x)$, $E_{MACRO}^o$ and $E_U^o$ by the
intersection between $y_C(x)$ and $y_I^l(x)$. Increasing $\theta$
there are new equilibria. For $\theta > \theta_{cf} =
y_c(0)-y_I(0)$ CF becomes at least LAS and $E_{micro}$ disappear.
On the right, as a consequence of the elementary properties of
continuous decreasing functions, increasing $\theta$ the equilibria
move and it is $x_{Eu}(\theta)>x_{Eu}(0)$,
$x_{E_{MACRO}}(\theta)<x_{E_{MACRO}}(0)$, and there
exists $\theta_r \in (0,y_I^r(x_{Eu})-y_C^r(x_{E_{MACRO}}))$ such that
for $\theta> \theta_r$ $E_{MACRO}$ and $E_U$ disappear.
Summarizing, when $\theta >
\tilde{\theta}=Max(\theta_{cf},\theta_r)$ then $CF$ is GAS (figure
\ref{lb}-subplot 3), because of proposition 4 of section
\ref{global}. If $\theta_r<\theta_{cf}$ then for
$\theta_r<\theta<\theta_{cf}$  $E_{micro}$ is GAS (figure
\ref{lb}-subplot 2), whereas when $\theta_{cf}<\theta_r$ for $
\theta_{cf}<\theta<\theta_r $ $CF$ is LAS and coexists with $E_U$
and $E_{MACRO}$(figure \ref{lbpat});

\item Aggressive tumors with variable sign $\Psi'(x)$.  In such a case, in the absence of therapy there may in the most complex case be one macroscopic equilibrium equilibrium: $E_{Macro}^o$ (GAS) and, of course, CF (unstable). Increasing $\theta$ two further equilibria may appear. The analysis is similar to the previous one (cfr figures \ref{lh} and \ref{lhpat}) and we may find a $\tilde{\theta}$ such that for $\theta>\tilde{\theta}$ CF is GAS. Note that when the tumor is aggressive
it is very likely that $\tilde{\theta}$ is "extremely high": $\tilde{\theta} >>
\sigma$;

\item Aggressive tumors with $\Psi'(x)<0$ \cite{[Ortega]}.  In such a case, in the absence of therapy there may in the worst case be one macroscopic equilibrium equilibrium: $E_{Macro}^o$ (GAS) and, of course, CF (unstable). Increasing $\theta$, if when $y_I(0)=y_c(0)$ it is $y_I'(0)<y'_C(0)$ then we may find two values $\theta_{cf}$ and $\tilde{\theta}>\theta_{cf}$ such that for $\theta_{cf}<\theta<\tilde{\theta}$ $CF$ is LAS and there is the birth of a third unstable equilibrium $E_u$. Finally for $\theta>\tilde{\theta}$ CF is GAS. Note that if when $y_I(0)=y_c(0)$ it is $y_I'(0)> y'_C(0)$ then $\theta_{cf} = \tilde{\theta}$.
\end{itemize}

When $f(0)=+\infty$ the total elimination cannot be achieved by immunotherapy alone.
Furthermore, even the suboptimal target of reducing the cancer to a microscopic size
in many relevant cases cannot be achieved for therapies of finite duration, however they may be long. In
fact, let it be $\Psi(x)>0$ (aggressive tumor) and let there be a unique GAS macroscopic equilibrium $E_{MACRO}$. By applying a CIT with $\theta$ sufficiently high there is a unique GAS microscopic equilibrium. However, when the therapy ceases $\theta$ falls to zero and the cancer restarts growing macroscopically, since $E_{MACRO}$ is again GAS. We note in brief that if the original equilibrium is microscopic (e.g. micrometastasis) the effect of the therapy is simply to create another and temporary microscopic equilibrium.

Let us suppose that there are three co-existing equilibria:
$E_{micro}^o$ (LAS), $E_U^o$ (Unstable and through which a separatrix $\Sigma^o$ passes) and $E_{MACRO}^o$ (LAS). Applying a CIT with $\theta > \tilde{\theta}$ there is an
unique GAS microscopic equilibrium. Thus at the end of the therapy (at $t=t_f$) depending on
the position of $P_f=(x(t_f),y(t_f))$ relatively to $\Sigma^o$, we have that either $(x(t),y(t))\rightarrow E_{micro}$ or $(x(t),y(t))\rightarrow E_{MACRO}$.

We note that $\theta$ acts a global bifurcation parameter, and we point out that these behavior may be observed in case of bounded $f(0)$ when therapy is applied for an insufficient time.

Finally, this simple analytical analysis may explain theoretically some numerical results of \cite{[Nani0]} on the relationships between the efficacy of the cure and the proliferation rate of cancer, and on the correlation between the burden of initial size and the probability of effectiveness of a therapy.

\subsection{Periodic Scheduling}
In the case of periodic drug schedulings, there is a periodically varying cancer-free solution $CF^*=(0,z(t))$, where $z(t)$ is the asymptotic periodic solution of:
\begin{equation}\label{eql}
 y' = -\Psi(0) y + \sigma  + \theta_m + \Omega(t)
\end{equation}
that, assuming $ \Omega(t) = \sum_{n=1}^{+\infty}{C_k Cos(k (2 \pi
/T) t -\zeta_n)}$, can be rewritten as::
\begin{equation}
 z(t) = \frac{\sigma + \theta_m}{\Psi(0)} + \sum_{n=1}^{+\infty}{ \frac{C_k}{\sqrt{\Psi^2(0)+k^2(\frac{2 \pi}{T})^2}} Cos(k \frac{2 \pi}{T}t-\zeta_n - Arg(\Psi(0)+ i k \frac{2 \pi}{T})  ) }.
\end{equation}
Note that if $T << 1 / \Psi(0) $ there is a filtering effect and $z(t)\approx (\sigma + \theta_m)/\Psi(0) $.

Two basic models of therapy may be:
\begin{itemize}
\item \begin{equation}\label{unr} \theta_u(t)= A (1 + b  cos(\omega t) ) \end{equation}
   which is rather unrealistic, but whose functional form is commonly used to assess the
   effect of periodic forcing on nonlinear systems. The asymptotic solution of (\ref{eql}) corresponding to (\ref{unr})
   is given by: $$z_u(t)= \frac{\sigma + A}{\Psi(0)} + \frac{A b}{\sqrt{\Psi^2(0)+\omega^2}} Cos(\omega t - Arg(\Psi(0)+ i \omega)  )   $$
\item the more realistic function: \begin{equation}\label{cos} \theta_{r}(t) = \frac{G}{1-Exp(-cT)}Exp(-c Mod(t,T)) \textrm{ , } \theta_m = \frac{G}{c T}, \end{equation}
which represent a boli-based delivery. The "shape" of
$\theta_r(t)$ depends on $c$ and the corresponding asymptotic
periodic solution of (\ref{eql}) is given by:
$$z_r(t)=\frac{\sigma}{\Psi(0)} + \frac{G}{\Psi(0)-c}\Big(
\frac{E^{-c Mod(t,T)}}{1-E^{-c T}}-\frac{E^{-\Psi(0)
Mod(t,T)}}{1-E^{-\Psi(0) T}}\Big)$$
\end{itemize}
In case of impulsive therapy, by solving the impulsive differential equation
\begin{equation}
 y' = -\Psi(0) y + \sigma \textrm{, } y(nT^+)=y(nT^-)+\gamma \textrm{, } n=0,1,\dots
\end{equation}
one obtains that:
\begin{equation}
 z(t) = \frac{\sigma}{\Psi(0)} + \frac{\gamma}{1-Exp(-\Psi(0)T)}Exp\Big(-\Psi(0)Mod(t,T)\Big).
\end{equation}
Furthermore, it is easy to show that the condition $ \sigma +
\theta_m > \sigma_{cr} $ guarantees the LAS of $CF$. In fact,
since the variational equations around $(0,z(t))$ are: $U'=(\alpha
f(0) - \phi(0)z(t) )U , W' =(\sigma q'(0)-\Psi'(0)z(t))U-\Psi(0)W
$, we obtain that $\alpha f(0) - \phi(0)<z(t)> <0 \Rightarrow U(t)
\rightarrow 0 \Rightarrow W(t) \rightarrow 0$, and since
$<z(t)>=(\sigma + \theta_m)/\Psi(0)$ we recover the LAS condition
$ \sigma + \theta_m > \sigma_{cr} $. Similarly, one may
demonstrate the GAS condition: $ \sigma + \theta_m > \sigma^* $.

\subsection{Numerical simulations}
We performed a set of simulations of immunotherapy on the basis of
the model proposed by Kuznetsov et al. \cite{[Kuznetsov]}, in
which: $$ \alpha f(x) = 1.636(1-0.002 x),\phi(x)=1 ,\beta(x)=
\frac{1.131 x}{20.19+x}, \sigma q(x) =
 0.1181 $$
$$ \mu(x)=0.00311 x + 0.3743 , $$
and
$$ t^{true}=9.9 t^{adim} days, (X,Y)= 10^6 (x,y) cells$$

We chose this model since its parameter values were fitted
from real data of chimeric mice \cite{[Kuznetsov]}. Note that the
dynamic of tumors in mouse is faster than that of human tumors,
and that for periods of about one day or less (i.e. $T< 0.101$) it
results that $ (1 /\mu(0)) >> T $. Moreover, $\mu'(x)=0.00311 <<1$ and
the tumor is not aggressive. We also performed simulations in a case of
a more aggressive tumor, for which we set $\mu(x)=10 (0.00311 x) +
0.3743$. For the non-aggressive tumor $\sigma_{cr} \approx 0.612$
and $ \sigma^* \approx 1.44 >> \sigma $.

It is worth noticing that in other kinds of anticancer therapies the shape of the therapy may be critical in determining whether or not the cancer will be eradicated \cite{[mbs]}.

In our simulations we assumed $\sigma +\theta_m > \sigma_{cr} $
which means that the mean value of the therapy, if given as CIT,
would enassure the LAS of the disease free equilibrium. Since for
each $T$ the mean value is constant, this means that in the limit
$c\rightarrow +\infty$ the therapy $\theta_r(t)$ tends to become
impulsive.

\begin{figure}[ht]
 \centering
 \includegraphics{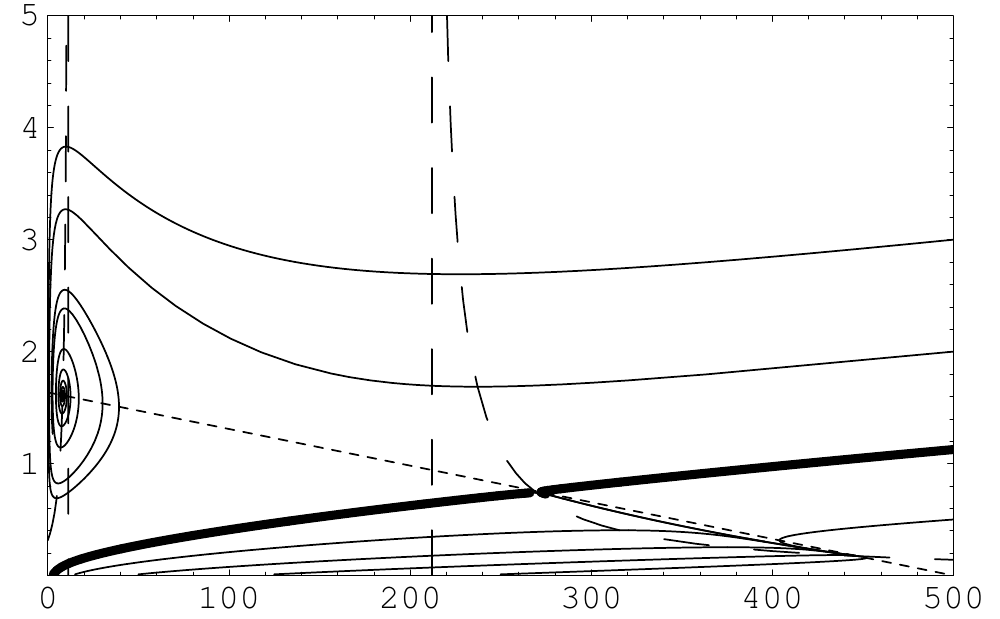}
  \caption{Non aggressive tumor: phase portrait of model \cite{[Kuznetsov]} in the absence of therapy. There are two LAS equilibria, whose basins of attraction are separated by the separatrix line (plotted with a thick line). The nullcline $ y_C(x)$ is plotted with short dashes, the nullcline $y_I(x)$ and its vertical asymptotes are plotted with long dashes.}
  \label{figure2}
\end{figure}

\begin{figure}[ht]
 \centering
 \includegraphics{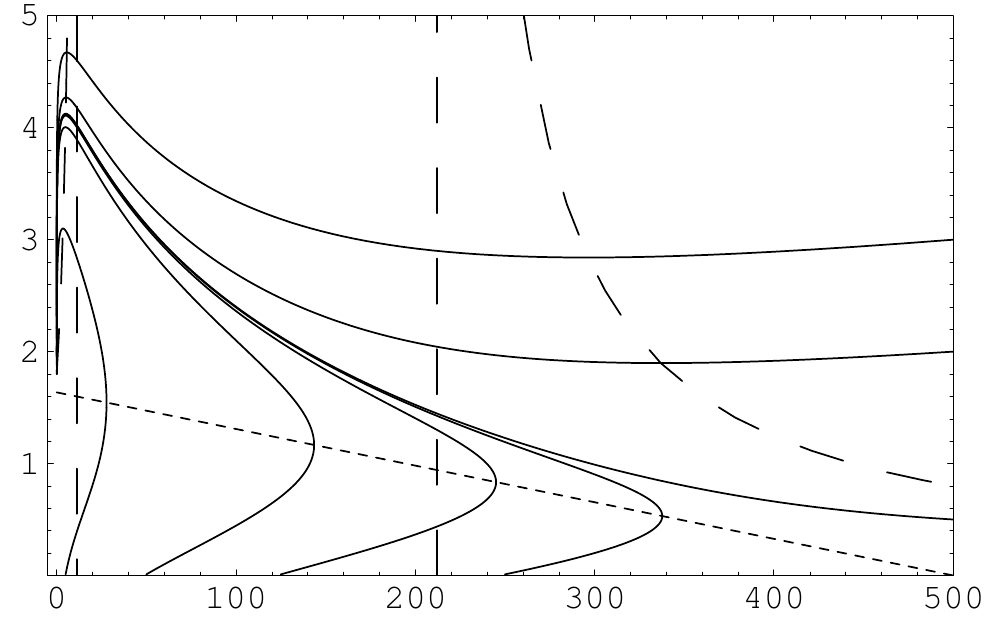}
  \caption{Non aggressive tumor: phase portrait of model \cite{[Kuznetsov]} in the presence of constant therapy with $\sigma + \theta_m =1.1 \sigma_{cr}$. There is a  tumor-free equilibrium $CF=(0,1.799)$, which is globally stable. The nullcline $ y_C(x)$ is plotted with short dashes, the nullcline $y_I(x)$ and its vertical asymptotes are plotted with long dashes. Note that the orbits stemming from initial points characterized by low $y(0)$ are characterized by an {\em initial } fast growth of the tumor size, followed by a regression to $0$. }
  \label{figure3}
\end{figure}

\begin{figure}[ht]
 \centering
 \includegraphics{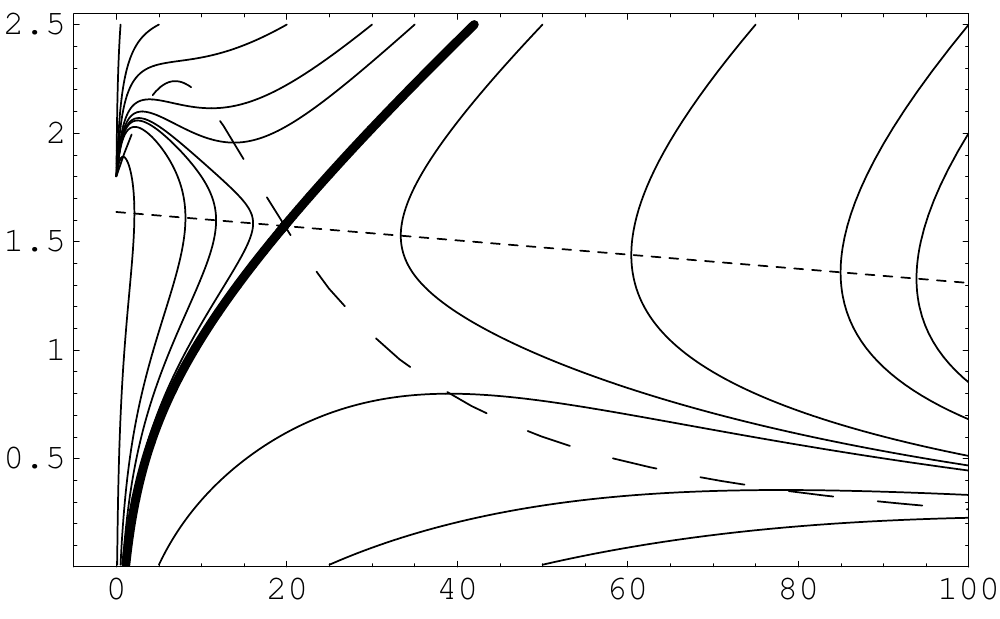}
  \caption{Aggressive tumor: phase portrait of model \cite{[Kuznetsov]} in presence of constant therapy. There is a  tumor-free equilibrium $CF=(0,1.799)$ and another LAS equilibrium, whose basins of attraction  are divided by a separatrix line (plotted with a thick line). The nullcline $ y_C(x)$ is plotted with short dashes, the nullcline $y_I(x)$ and its vertical asymptotes are plotted with long dashes.}
  \label{kuzzoomCT}
\end{figure}

\begin{figure}[ht]
 \centering
 \includegraphics{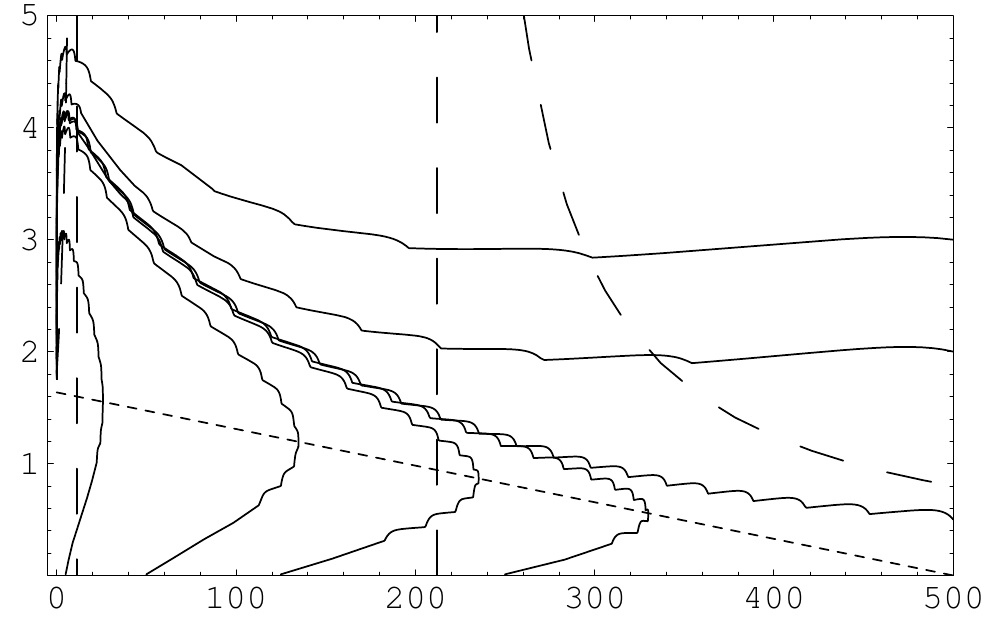}
  \caption{Non aggressive tumor: phase portrait of model \cite{[Kuznetsov]} in presence of periodic therapy $\theta_r(t)$ with $T=0.202$ (=2 days) and $1 / c =0.1 T$. There is a tumor free equilibrium $(0,z(t)) \approx (0,1.799) $ which remains GAS. The nullcline $ y_C(x)$ is plotted with short dashing, the nullcline $y_I(x)$ and its vertical asymptotes are plotted with long dashes.}
  \label{napt}
\end{figure}

\begin{figure}[ht]
 \centering
 \includegraphics{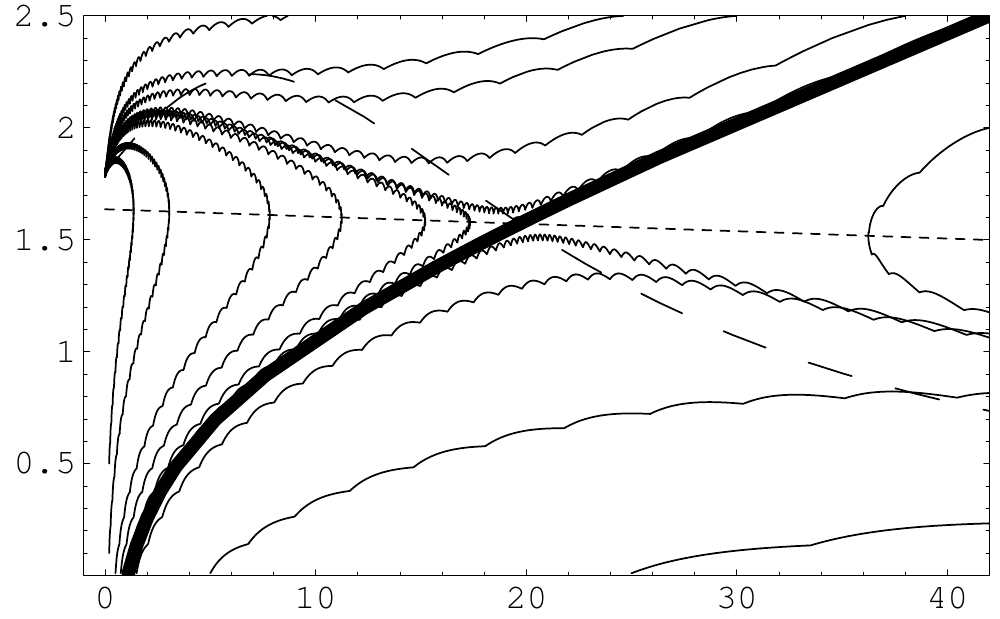}
  \caption{Aggressive tumor: phase portrait of the model \cite{[Kuznetsov]} in the presence of periodic therapy with $T=0.202$ (=2 days) and $1 / c =0.5 T$. The basins of attraction of the tumor-free equilibrium $CF^* = (0,z(t)) \approx (0,1.799) $ and of the macroscopic size equilibrium remain near unchanged respect to the CIT scheduling (the basin of CF is slightly greater than in the CIT). The nullcline $ y_C(x)$ is plotted with short dashing, the nullcline $y_I(x)$ and its vertical asymptotes are plotted with long dashes.}
  \label{kuzzoomPTa}
\end{figure}

\begin{figure}[ht]
 \centering
 \includegraphics[width=5cm,height=5cm]{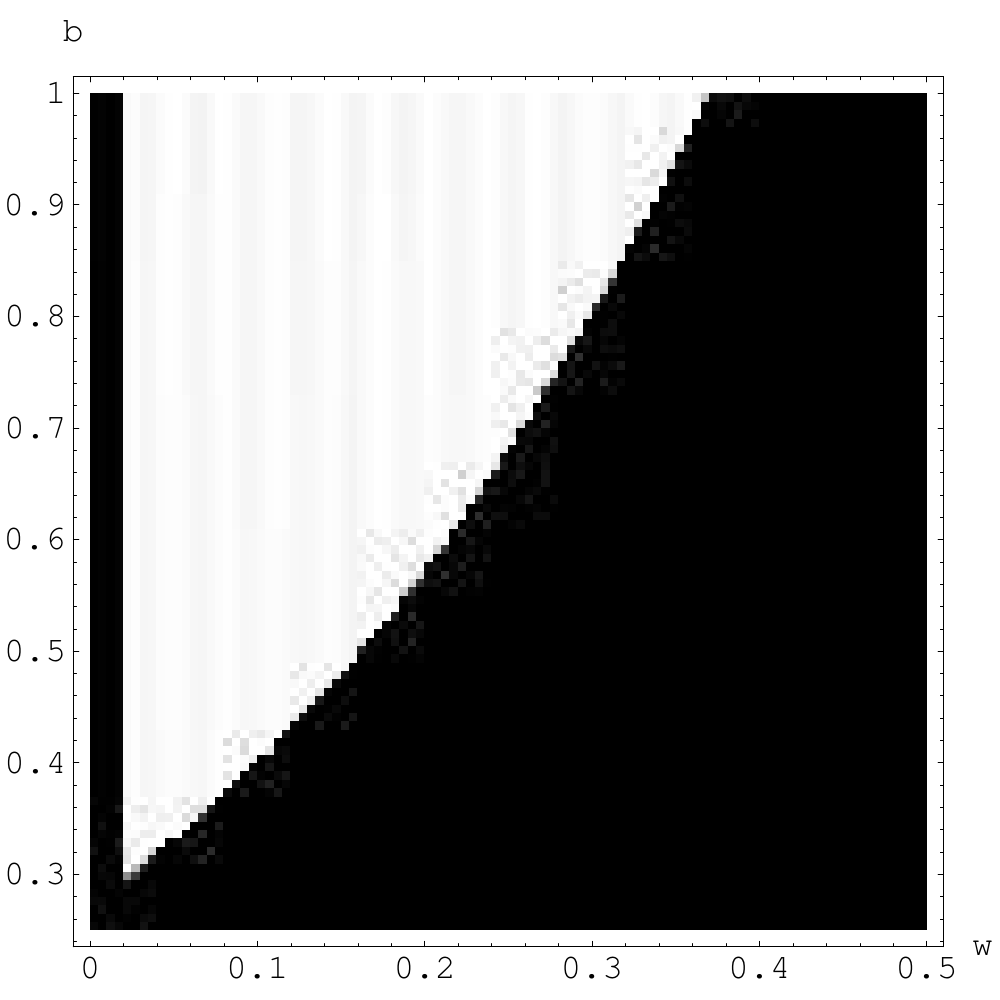}
 \includegraphics[width=5cm,height=5cm]{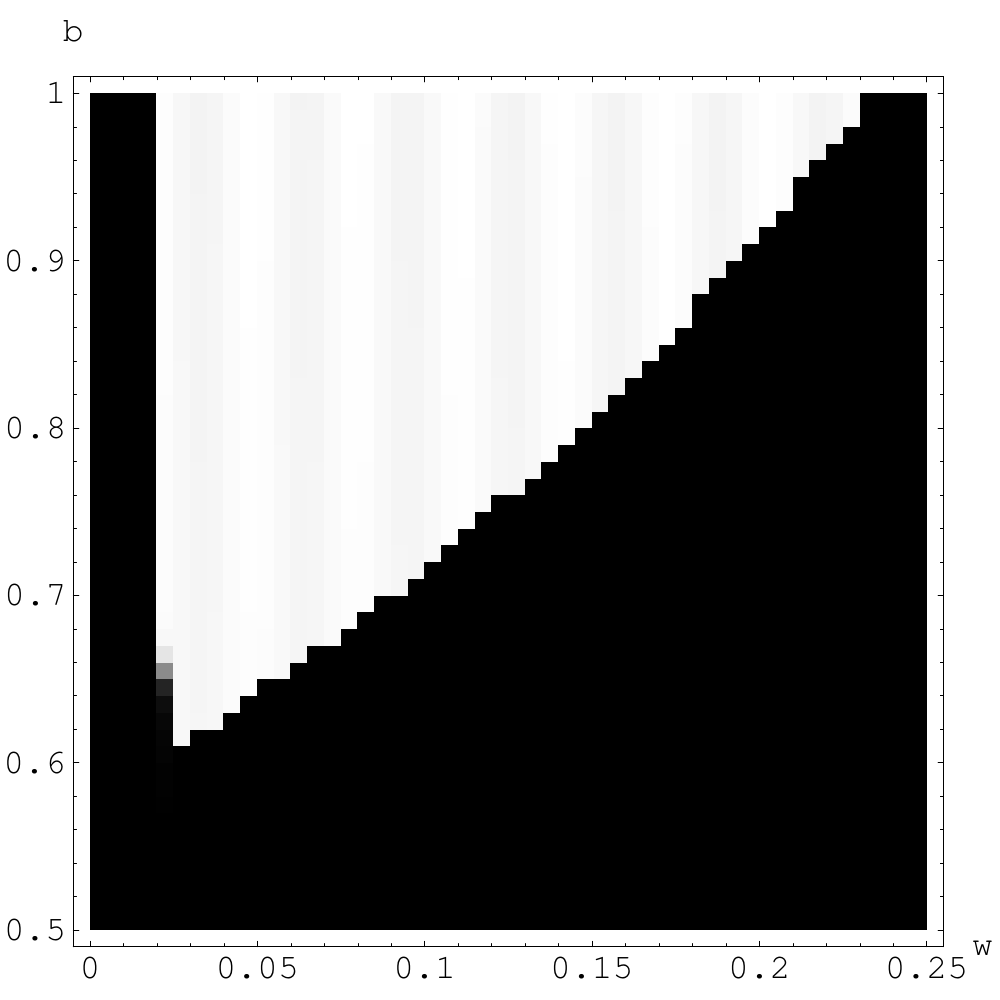}
 \includegraphics[width=5cm,height=5cm]{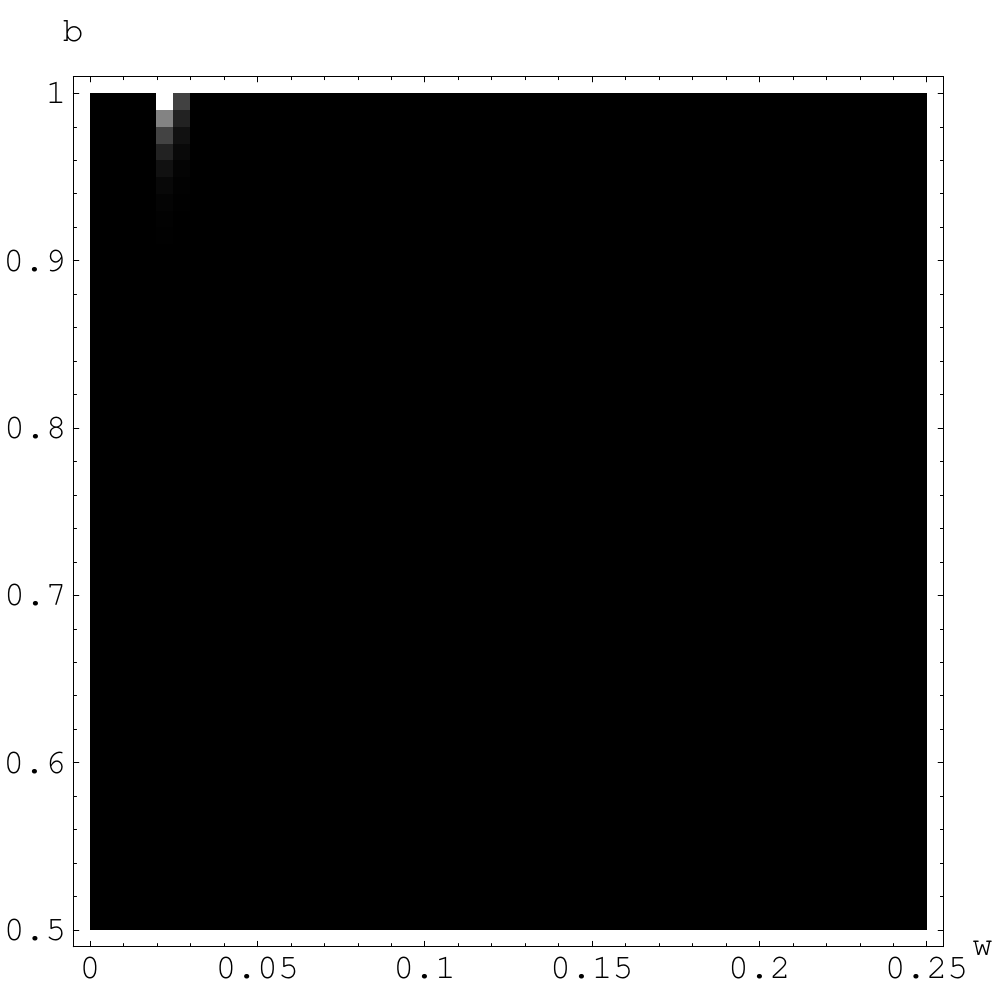}
  \caption{Aggressive tumor in the presence of immunotherapy $\theta_u(t)$: growth behavior in function of the parameters $(\omega,b)$. Black points correspond to eradication, white points to macroscopic growth. The initial condition for all is $(40,2.5)$ (chosen near the separatrix for the constant therapy). Left: $\sigma + A = 1.01 \sigma_{cr}$, central: $\sigma + A = 1.1 \sigma_{cr}$ and right: $\sigma + A = 1.25 \sigma_{cr}$. Note that the frequencies which do not allow eradication are very low, corresponding to absolutely unrealistic periods for the therapy.}
  \label{bifubw}
\end{figure}

We found that:
\begin{itemize}
    \item In the absence of therapy:  non-aggressive tumor has two stable equilibria: one slightly less than the carrying capacity and the other corresponding to a small tumor (see phase portrait in figure \ref{figure2}). For the highly aggressive tumor there is one GAS equilibrium slightly less than the carrying capacity;
     \item With constant therapy:  the non-aggressive tumor has a  cancer-free equilibrium, which results to be GAS  (figure \ref{figure3}). Note that the orbits stemming from initial points characterized by low values of the number of immune system cells are characterized by an {\em initial } rapid growth of the tumor size, followed by a regression to $0$. Biologically, the therapy might seem to help the tumor growth, instead of fighting it. For the highly aggressive tumor, the cancer free equilibrium is LAS, but there is also a high size LAS equilibrium (\ref{kuzzoomCT});
     \item In the presence of periodic therapy with $\theta_r(t)$, for both types of tumors the phase portrait is roughly similar to that of the constant therapy: the cancer-free periodic solution remains GAS for the non aggressive tumor (figure \ref{napt}). For the aggressive tumor there is the coexistence of the cancer free solution with a solution fluctuating around high values of the cancer size (near the equilibrium of the constant therapy). The two basins of attraction for the aggressive tumor remain unvaried with respect to those of the constant therapy (figure \ref{kuzzoomPTa}).
     \item For $\theta_r(t)$ the dependence  of the qualitative properties of the system on the parameter $c$ is not critical.
     \item For aggressive tumor and $\theta_u(t)$, it may occur that, given an initial point, the eradication is also a function of parameters $b$ and $\omega$, but this happens only for unrealistically high values of the therapy period (figure \ref{bifubw}), e.g. $T>100$ days. These results may be roughly explained considering that for $ T >> Max(1/\Psi(0) ,1 / \alpha) $, one may approximately consider $\theta_u(t)$ as  constant;
         \item Both with CIT and with periodic therapy $y(t)$ may reach values considerably higher than the physiological value $\sigma/\mu(0)$, which might model some serious side effects of immunotherapies due to the excess of immunocompetent cells \cite{[PPV],[VHR]}.
\end{itemize}

\begin{figure}[ht]
 \centering
 \includegraphics{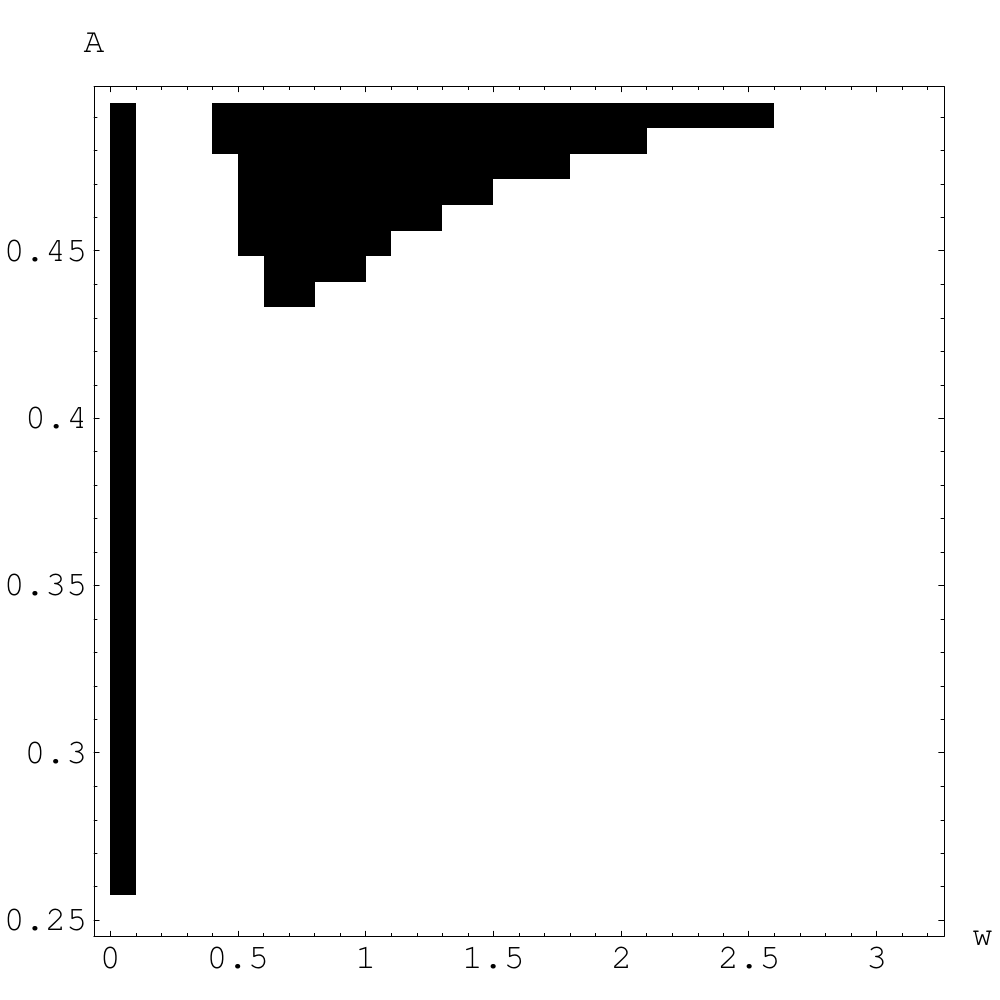}
  \caption{Aggressive tumor in the presence of immunotherapy $\theta_u(t)$: growth behavior in function of the parameters $(\omega,A)$ for $b=1$, with $A<\sigma_{cr}-\sigma$. Black points correspond to eradication, white points to macroscopic growth. The initial condition for all is $(40,2.5)$ (chosen near the separatrix for the constant therapy).}
  \label{bifuaw}
\end{figure}

For the sake of completeness, we also performed some simulations in
which $0<A< \sigma_{cr} - \sigma$ and for which there were high oscillations
($b=1$). We obtained the result that for low frequencies, there may be points
in the $(\omega,A)$ plane for which eradication is possible. See fig. (\ref{bifuaw}).

\begin{figure}[ht]
 \centering
 \includegraphics{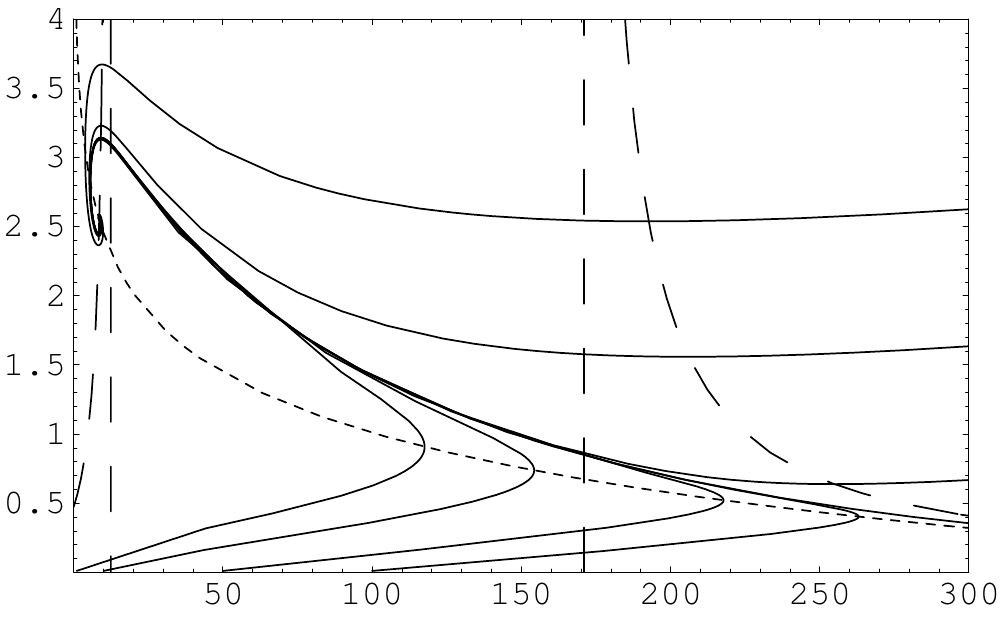}
  \caption{Simulation of the modified Kutnetsov model with CIT and $\theta_m = 0.5 \sigma $.}
  \label{c3}
\end{figure}

\begin{figure}[ht]
 \centering
 \includegraphics{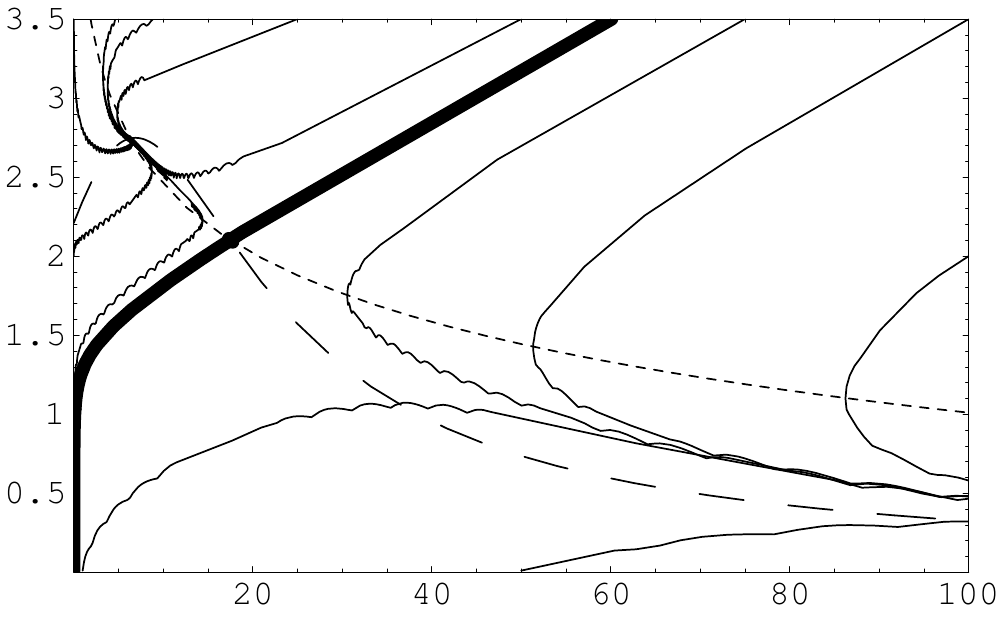}
  \caption{Simulation of the modified Kutnetsov model with periodic therapy, $T= 0.202$ (2 days), $1/c = 0.5 T$ and $\theta_m = 6 \sigma$.}
  \label{c6}
\end{figure}

Finally, we performed simulations for a hybrid model similar to
that by Kuznetsov et al. \cite{[Kuznetsov]}, but in which we
assumed: $$ \alpha = 0.626, f(x) = Log(\frac{500}{x}),$$ the other
parameters being as before. We choose the value $\alpha = 0.626$
in order to minimize the difference with $f(x)$ in
\cite{[Kuznetsov]}. The results of the simulations are very close
to those relative to the logistic case: figures \ref{c3}, \ref{c6}. In order to obtain via CIT
the reduction to the microscopic state $\theta>8.4 \sigma$ about
is required.

The analytical and numerical results obtained
in this section may be usefully compared with two similar works of
the recent literature which focus on Adoptive Cellular
Immunotherapy. An excellent analytical work is \cite{[NaniMbs]},
who, however, cannot be fully compared with our results because it
refers to tumors which have no action in stimulating immune cells.
Furthermore, its formulae for the global stability of the
cancer free equilibrium are not expressed as a function of the
parameters of the therapy. In a very interesting paper \cite{[Kirschner]} some results
similar to ours are obtained through numerical bifurcations on a
three dimensional model in which the direct immunogenicity of
tumors is expressed as an additive term $c x$.  As previously
stressed, in the absence of therapy and of influx of immunocompetent
cells both our model and the model in \cite{[Kirschner]} show the possibility
of having periodic solution, which in \cite{[Kirschner]} are shown
to be present also in some cases in which there is therapy. We
notice in brief that a term $c x$ may be formally embedded in our
generic function $\sigma(x)$.

\section{Concluding Remarks}
It is interesting to use well established conceptual frameworks of
ecological models to model competition phenomena in human biology,
but it is important to pay attention to the whole ecological
modeling aspect, such as the basic requirement of the positivity
of the solutions. Even if model \cite{[Sotolongo]} violates the
positivity rule, it is valuable because it may be read as a model
which takes into account a disease-induced depression in the
influx of lymphocytes. Then, instead of proposing another specific
model, we preferred to add this new feature to a family of
equations, and to analyze its properties. We stressed also
that models which do not allow the possibility to have LAS
tumor-free solutions should be cautiously considered. The general
family (\ref{GenFamilyX})-(\ref{GenFamilyY}) may be, of course,
further generalized following Volterra's ecological theory,
i.e. by considering that there may be a delay between the
consumption of a prey and the birth of a predator (see also
\cite{[Nani0],[Galach],[Suzanna]}), i.e. by allowing a delay
$\tau$ with probability density $\varrho(\tau)$. This delayed
model and stochastic models will be the subject of further
investigations.

Finally, we would like to illustrate some qualitative medical
inferences from the investigations that we have here proposed. The
main problem of immunotherapy is that, as it is clear from our
analysis and simulations, in general, eradication may be
possible but is dependent on the initial conditions $(x(0),y(0))$.
However, the IC are in medical practice unknown or known with
very large confidence intervals (cfr.  \cite{[Dwery]} for the
cancer cells at the start of a radiotherapy and). This makes it
impossible to plan an anticancer therapy based solely on this
therapy. This is a peculiarity of immunotherapy, since there are
other kinds of anticancer cures for which a globally stable
eradication is possible \cite{[mbs]}. However, in our simulations
we have seen that in some particular cases the model
\cite{[Kuznetsov]} predicts that globally stable eradication is
possible also in case of immunotherapy, but that it depends on the
"degree of aggressiveness" of the cancer, i.e., on the framework
of the model \cite{[Kuznetsov]}, on the parameter $\mu_1$.
However, $\mu_1$ is difficult to be estimated (as a range) and, in
particular, on single patients. If in the future it might be
possible, the option to use immunotherapy as main strategy, for
relatively small "non aggressive" tumors, could be seriously
considered. Furthermore, we showed that the behavior of the system
does not depend on the amplitude of fluctuations of $\theta(t)$,
so that the option of continuous intravenous infusion is not,
dynamically, better than the boli based therapy. This result may
be of interest, since continuous intravenous infusion may cause
major practical problems to the patients. Finally, in case of
disease aggressive towards the immune system, since our
simulations indicated that all the positive quadrant is GAS
towards a macroscopic disease in absence of therapy and low
$\sigma$, whereas in the presence of therapy the eradication is
possible in an adequate basin (see figure \ref{kuzzoomCT}), we may
infer that a conventional therapy should be followed by
immunotherapy to increase the probability of total remission.
\section{Acknowledgements} I am very grateful to two anonymous referees who helped me to improve greatly this paper. Special thanks to Prof. Alberto Gandolfi who read the drafts of this papers and gave me precious suggestions, and, for their precious bibliographical help, to Giorgio "Leppie" Donnini, to William Russell-Edu Esq. and to Matteo "Furjo" Sisa.

\end{document}